\documentclass[aps,prx,twocolumn, groupedaddress, nobalancelastpage]{revtex4-2}

\usepackage{blindtext,amssymb,gensymb,siunitx,textgreek,textcomp,mathtools,bm,soul}
\usepackage{wasysym} 
\usepackage[version=3]{mhchem} 
\newcommand\upe{\mathord{\mathrm{e}}}

\usepackage[utf8]{inputenc}
\usepackage[T1]{fontenc}
\usepackage [english]{babel}
\usepackage [autostyle, english = american]{csquotes}
\MakeOuterQuote{"}
\usepackage{subfiles}
\usepackage{fancyvrb}

\usepackage[margin=0.625in]{geometry}

\usepackage{caption,graphicx}
\usepackage{tablefootnote}

\usepackage{booktabs}
\usepackage[table,xcdraw]{xcolor}
\usepackage[hidelinks]{hyperref}
\usepackage{cleveref,xr}

\begin{document}


\title{The role of disorder in the synthesis of metastable zinc zirconium nitrides}

\author{Rachel Woods-Robinson\textsuperscript{*,1,2,3}, Vladan Stevanović\textsuperscript{4,3}, Stephan Lany\textsuperscript{3}, Karen N. Heinselman\textsuperscript{3}, Matthew K. Horton\textsuperscript{2}, Kristin A. Persson\textsuperscript{5,2}, Andriy Zakutayev\textsuperscript{*,3}}

\affiliation{\textsuperscript{1}Applied Science and Technology Graduate Group, University of California at Berkeley, Berkeley, CA, 94720 USA, \textsuperscript{2}Materials Sciences Division, Lawrence Berkeley National Laboratory, Berkeley, CA, 94720 USA, \textsuperscript{3}Materials Science Center, National Renewable Energy Laboratory, Golden, Colorado, 80401 USA \textsuperscript{4}Department of Metallurgical and Materials Engineering, Colorado School of Mines, Golden, Colorado, 80401 USA,
\textsuperscript{5}Department of Materials Science and Engineering, University of California at Berkeley, Berkeley, CA, 94720 USA}

\date{\today}


\begin{abstract}

In materials science, it is often assumed that ground state crystal structures predicted by density functional theory are the easiest polymorphs to synthesize. Ternary nitride materials, with many possible metastable polymorphs, provide a rich materials space to study what influences thermodynamic stability and polymorph synthesizability. For example, \ce{ZnZrN2} is theoretically predicted at zero Kelvin to have an unusual layered "wurtsalt" ground state crystal structure with compelling optoelectronic properties, but it is unknown whether this structure can be realized experimentally under practical synthesis conditions. Here, we use combinatorial sputtering to synthesize hundreds of \ce{Zn_{$x$}Zr_{1-$x$}N_{$y$}} thin film samples, and find metastable rocksalt-derived or boron-nitride-derived structures rather than the predicted wurtsalt structure. Using a statistical polymorph sampler approach, it is demonstrated that although rocksalt is the least stable polymorph at zero Kelvin, it becomes the most stable polymorph at high effective temperatures similar to those achieved using this sputter deposition method, and thus corroborates experimental results. Additional calculations show that this destabilization of the wurtsalt polymorph is due to configurational entropic and enthalpic effects, and that vibrational contributions are negligible. Specifically, rocksalt- and boron-nitride-derived structures become the most stable polymorphs in the presence of disorder because of higher tolerances to cation cross-substitution and off-stoichiometry than the wurtsalt structure. This understanding of the role of disorder tolerance in the synthesis of competing polymorphs can enable more accurate predictions of synthesizable crystal structures and their achievable material properties.

\end{abstract}

\maketitle

\section{Introduction}

Computational materials discovery is a rapidly progressing research field, with the potential to revolutionize how materials are designed and developed. However, determining whether a given predicted crystalline material is \textit{actually experimentally synthesizable} remains a key challenge. One common assumption in computational materials research is that the ground state structure predicted by density functional theory (DFT) within the zero temperature (0 K) approximation, or structures with energies near the ground state energy, are the most likely to be experimentally realized. Conversely, another assumption is that increased energetic instability (i.e. formation energy farther away from the ground state energy) correlates with an increased difficulty to synthesize. However, neither of these assumptions necessarily hold, as demonstrated by multiple experimental and computational studies.\cite{sun2017thermodynamic} Recent work has emerged to further explore synthesizability in metastable materials,\cite{stevanovic2016sampling, sun2016thermodynamic, aykol2018thermodynamic, aykol2019network} but so far computational materials researchers still cannot confidently answer the following question: "can this predicted material be synthesized?"\cite{horton2021perils} Thus, as materials databases grow and structure predictions yield new predicted compounds for high-throughput screenings, it is increasingly pertinent that the computational materials discovery community develops comprehensive methods for assessing synthesizability so that misleading false positives and negatives can be avoided.

Nitrides provide a compelling class of materials through which to examine synthesizability, in part because they are more likely than any other anion class to crystallize in metastable phases.\cite{kroll2003pathways, sun2016thermodynamic, aykol2018thermodynamic, greenaway2020ternary} Recent computational predictions have yielded a multitude of new ternary nitride materials to explore,\cite{hinuma_discovery_2016, sun2019map} yet an understanding of which polymorphs are experimentally synthesizable remains elusive. The chemical and structural richness of this emerging class of materials, including their mixed ionic-covalent nature compared to oxides, provides new candidates for various applications such as hydrogen storage, photovoltaic (PV) devices, and light-emitting diodes (LEDs). One such emerging class of ternary nitrides is the II-IV-\ce{N2} family, ternary analogs of GaN and promising candidate for PV absorbers and green LEDs. II-IV-\ce{N2} materials are commonly studied in two prototype classes: (1) wurtzite-derived (WZ) structures, with four-fold coordinated cations (e.g. Zn-based \ce{ZnSnN2}, \ce{ZnGeN2}, \ce{ZnSnP2}, \ce{ZnSiP2}),\cite{martinez2017synthesis} and (2) rocksalt-derived structures (RS), with six-fold coordinated cations (e.g. Mg\textit{TM}\ce{N2}).\cite{bauers2019ternary} Some compounds (e.g. \ce{MgSnN2}) have been shown to co-crystallize in both of these configurations at certain growth conditions, such as at ambient temperature at Mg-rich stoichiometries or at increased synthesis temperature on GaN substrates.\cite{kawamura2020synthesis, greenaway2020combinatorial} However, these two structure classes are just a small subset of possible structure classes in the rich space of ternary nitrides; it remains unknown which other II-IV-\ce{N2} polymorph structures and chemistries are stabilizable.

\begin{figure}
    \centering
    \includegraphics[width=80mm]{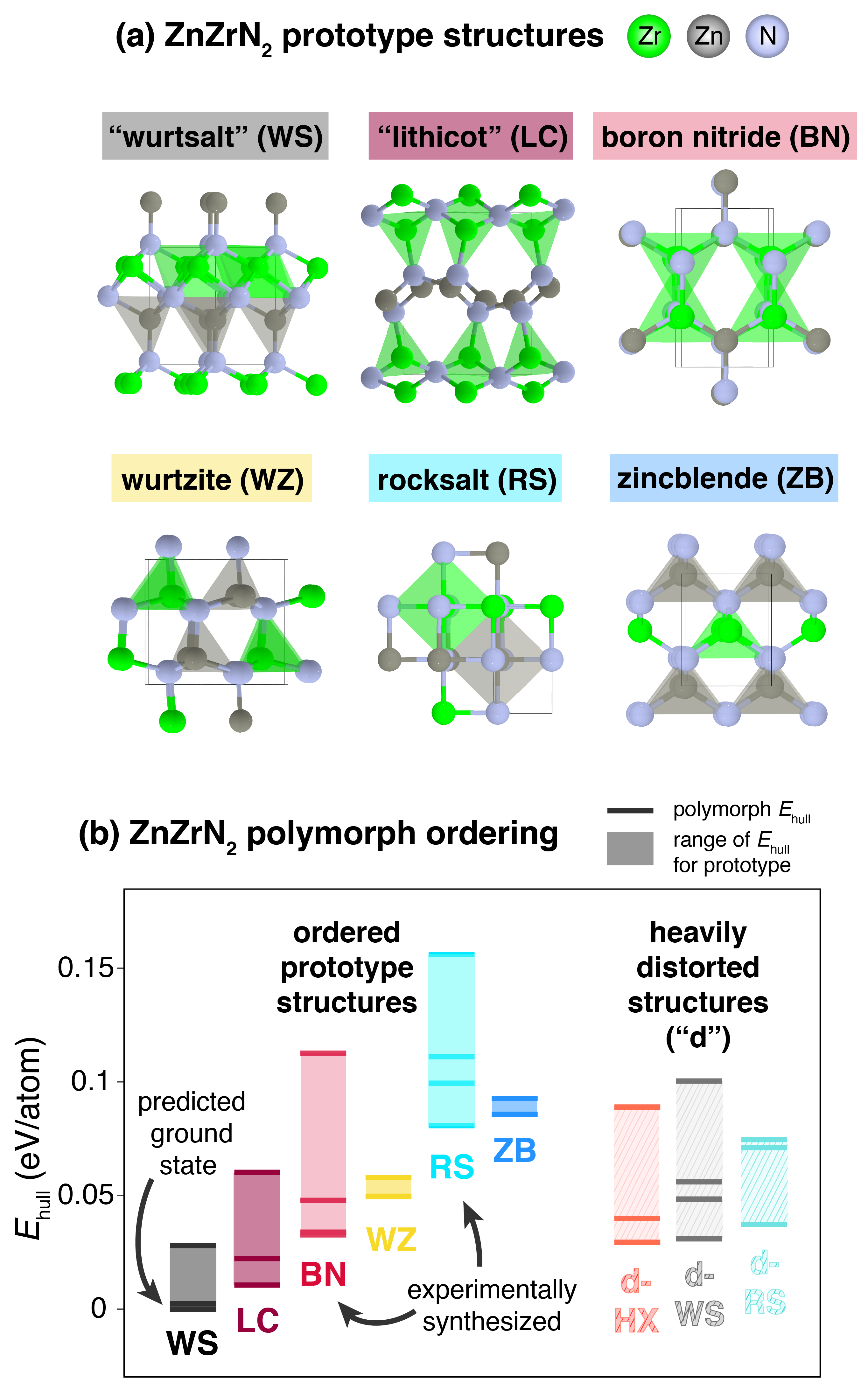}
    \caption{(a) Representative crystal structures for each of the six ordered structure prototype classes, and (b) energy ordering of the predicted ordered polymorphs, grouped by structure prototype class, with labeled experimentally synthesized phases from this study. Horizontal solid lines in (b) correspond to calculated $E_\mathrm{hull}$ values of individual polymorphs, and shaded regions correspond to the range of $E_\mathrm{hull}$ for a given prototype class. Heavily distorted versions of the prototype structures, designated with a "d-" prefix, are plotted separately on the right of (b), and "HX" designates distorted hexagonal structures as described in the text.}
    \label{fig:prototypes}
\end{figure}

Of particular interest to this study is the experimentally empty region of phase space in the zinc zirconium nitride (Zn-Zr-N) material system, in particular at its II-IV-\ce{N2} composition \ce{ZnZrN2}, which serves as a case study to gain insight for ternary nitrides as a whole. In contrast to other II-IV-\ce{N2} materials, \ce{ZnZrN2} (as well as isoelectronic \ce{ZnHfN2}) has a DFT-predicted $P3m1$ (156) space group ground state structure---a layer of Zn atoms tetrahedrally coordinated by N (wurtzite-like), a layer of Zr atoms octahedrally coordinated by N (rocksalt-like), and alternating Zn and Zr layers---which has been corroborated by three different computational studies using three distinct structure prediction algorithms with DFT relaxations.\cite{hinuma2016discovery, tholander2016strong, sun2019map} This structure is analogous to sulfosalt \ce{ScCuS2}, though a corresponding mineral name could not be located;\cite{dismukes1971physical, scanlon2010stability} thus, we herein refer to this structure type as "wurtsalt" (WS), an amalgam of \textit{wurt}-zite and rock-\textit{salt}, and depict the \ce{ZnZrN2} WS structure in the top left of \textbf{\autoref{fig:prototypes}}(a), alongside other polymorphs. Despite these predictions, no semiconducting nitride materials in the \ce{Zn_{$x$}Zr_{1-$x$}N_{$y$}} ternary space have ever been stabilized experimentally, and it has not yet been investigated whether any other polymorphs exist. 

In this study, we demonstrate that certain polymorphs can be preferentially stabilized or destabilized due to their tolerance to disorder. First, a set of 28 possible \ce{ZnZrN2} polymorphs are predicted and investigated computationally. Next, combinatorial sputter synthesis is used to explore the full cation phase space of \ce{Zn_{$x$}Zr_{1-$x$}N_{$y$}} heterovalent heterostructural alloys (note that for simplicity, "alloy" will be used herein to represent this system), focusing on the region where $y \approx$ 1 and $x \approx$ 0.5 close to the \ce{ZnZrN2} stoichiometry. Under this range of experimental sputtering conditions, the cation-ordered WS ground state structure predicted by DFT at 0 K is \textit{not} synthesized. Instead, a disordered metastable rocksalt (RS) polymorph is synthesized close to the \ce{ZnZrN2} stoichiometry, and a metastable hexagonal boron-nitride-derived (BN) phase is observed at higher Zn concentrations ($x \gtrsim$ 0.5). We note that the term “metastable” herein refers to solids that are metastable with respect to computed DFT energies at 0 K, as described in Sun et al.\cite{sun2016thermodynamic} To understand the effects of disorder on the synthesis of \ce{ZnZrN2}, a series of computational methods are then applied. We start with the 0 K DFT approximation to reflect how Zn-Zr-N polymorphs would be energetically ranked in computational databases, and then take into account configurational entropy and enthalpy to account for temperature, as well as off-stoichiometry. These \ce{ZnZrN2} results suggest that it is necessary to consider the effects of disorder tolerance on energetic stabilization in possible polymorphs when investigating other new ternary nitrides, and new computationally predicted materials in general.

\section{Methods}

\subsection{Synthesis}

Thin film samples were grown using radio frequency (RF) co-sputter deposition and the combinatorial method, with a total of 24 thin film combinatorial ``libraries'' of \ce{Zn_{$x$}Zr_{1-$x$}N} deposited on 50 $\times$ 50 mm fused silica substrates in two different sputter chambers, with various experimental conditions. After exploratory depositions, 11 libraries are reported here, all grown in the same chamber. The chamber setup consists of 2 inch precursor sputter targets of metallic Zr and Zn, with sputter guns pointed towards one another to result in a gradient in cation composition, as depicted in the SM. The samples reported in this study are deposited in an Ar/\ce{N2} environment, with a chamber base pressure of $\sim\num{2e-7}$ Torr, growth pressure of 5 mTorr, and gas flow rates of Ar and \ce{N2} both of 6 sccm. In an attempt to increase nitrogen chemical potential, a nitrogen cracker is operated during growth at 300 W with 0 W reflected power, as described elsewhere. RF power is varied from 30--100 W on the gun with the Zn target, and 20--100 W on the gun with the Zr. Temperature gradient methodology and associated temperature calibrations are described elsewhere\cite{subramaniyan2014non, fioretti2015combinatorial} and in the SM.

\subsection{Characterization}

Material composition and structure was characterized with customized combinatorial measurement tools, as described elsewhere, with a 4$\times$11 mapping grid projected onto each sample library resulting in 44 data points per library and thus $>$400 unique compositional data points total in this study. Analysis is conducted using the customized COMBIgor software package.\cite{talley2019combigor}. Film cation composition, i.e. $x$ in \ce{Zn_{$x$}Zr_{1-$x$}N}, and film thickness were determined using mapping style X-ray fluorescence (XRF) spectroscopy and Dektak profilometry. Since nitrogen (as well as spurious oxygen) cannot be resolved with XRF, a select number of samples are measured using Rutherford backscattering spectrometry (RBS) to confirm cation composition and to measure the anion content in films. RBS was performed at NREL on a National Electrostatics Corporation 3S-MR10 instrument with a 2 MeV alpha particle beam at a current of 70 nA. The total accumulated charge was 320 $\mu$C, and the RBS detector was mounted in a backscatter configuration at 140$^{\circ}$. Analysis was performed with the RUMP package. Structural analysis mapping was performed for all libraries with X-ray diffraction (XRD) on a Bruker D8 Discover with a $\theta$–2$\theta$ geometry, Cu K$\alpha$ radiation, and a proportional 2D detector. Measurements are complimented for 11 libraries of interest at Beam Line 1-5 at the Stanford Synchrotron Radiation Lightsource (SSRL) with Wide Angle X-ray Scattering (WAXS). 2D scattering was collected with a Rayonix 165 CCD Camera at grazing incidence at an incident energy of 12.7 keV.

\subsection{Polymorph structure generation}

Candidate ordered polymorphs were generated using kinetically limited minimization (KLM)\cite{sharan2021computational} and ionic substitution of prototypes from other ternary nitrides.\cite{greenaway2020combinatorial} Unique structures that emerged from the polymorph sampler random structure searching were also included as ordered polymorphs.\cite{stevanovic2016sampling} To create a set of ordered prototype "alloys" across the ZrN--ZnN tieline (i.e. \ce{Zn_{$x$}Zr_{1-$x$}N}), we performed cation substitution in each of the ordered \ce{ZnZrN2} polymorph structures where $y$ = 1 and $x$ = 0, 0.25, 0.50, and 0.75. Details and structure matching are described in the SM.

To account for configurational degrees of freedom and associated entropic contributions to free energy, the “polymorph sampler” statistical approach of Stevanović et al.\cite{stevanovic2016sampling, jones2017polymorphism, jones2020glassy} was modified to include cation lattice disorder in the \ce{ZnZrN2} system. The modification pertains mainly to structure classification and the statistical treatment (see SM). Using random structure sampling, we generated a set of 5,000 random superlattice (RSL) \ce{ZnZrN2} structures with 24-atom cells.

\ce{ZnZrN2} structures approximating random disorder were simulated using the special quasirandom structure (SQS) method, which models random atomic decorations on a lattice in unit cells larger than most ordered structures but small enough to converge reliably with DFT.\cite{zunger1990special} This is achieved by searching for unit cells that reproduce or approximate pair (or higher order) correlation functions by minimizing an objective function (see SM). We calculated a set of SQS structures with 64 atoms for each \ce{ZnZrN2} structure class using the \texttt{ATAT} package,\cite{van2009multicomponent, van2013efficient, van2019alloy} selecting only structures with the lowest objective functions. Each SQS structure was assigned to its closest structure prototype class via a structure-matching algorithm to account for any SQS structures that may have relaxed to a different geometry.

\subsection{First principles calculations}

Density functional theory (DFT) calculations were performed using the projector augmented wave (PAW) method\cite{blochl1994projector, kresse1999ultrasoft} as implemented in the Vienna \textit{Ab Initio} Simulation Package (VASP)\cite{kresse1993ab, kresse1996efficient}, first within the Perdew-Burke-Enzerhof (PBE) Generalized Gradient Approximation (GGA) formulation of the exchange-correlation functional.\cite{perdew1996generalized} Cutoff, convergence, and correction criteria are described elsewhere.\cite{ONG2013314, jain2013commentary} To estimate energetic contributions from vibrational degrees of freedom for structures of interest, density functional perturbation theory (DFPT) calculations for gamma ($\Gamma$) point phonons ($q$ = 0) are run on representative polymorphs (see SM for details).

The ordered \ce{ZnZrN2} polymorph structures, 64-atom SQS structures, and alloy calculations ($x$ = 0, 0.25, 0.5, 0.75), were relaxed first with a PBE functional, then with PBE using a Hubbard U correction ("PBE+U") of 3 eV/atom for Zr as benchmarked by Stevanović and coworkers,\cite{stevanovic2012correcting} and also with the SCAN meta-GGA functional, which has been demonstrated to more accurately predict polymorph orderings with the trade-off of a higher computational cost.\cite{sun2015strongly, yang2019rationalizing, greenaway2020combinatorial} The SCAN results are reported for all calculations herein, except for DFPT which uses PBE+U and the polymorph sampler structures which were relaxed using PBE+U since SCAN is too computationally expensive for 5,000 structures. Additional calculation details are reported in the SM.

\section{Results}

\begin{table}[]
\caption{Representative ordered polymorphs from each prototype class with the lowest $E_\mathrm{hull}$ (see SM for full list of polymorphs and energies)}
\label{tab:polymorph_energies}
\resizebox{90mm}{!}{%
\begin{tabular}{@{}cccccccc@{}}
\toprule
\rowcolor[HTML]{EFEFEF} 
    \textbf{\begin{tabular}[c]{@{}c@{}} Prototype \\ class, $\bm{k}$\end{tabular}} & 
    \textbf{\begin{tabular}[c]{@{}c@{}} Space \\ group\end{tabular}} & 
    \textbf{\begin{tabular}[c]{@{}c@{}}$\bm{\#}$ of \\ atoms\textsuperscript{†}\end{tabular}} &
    \textbf{\begin{tabular}[c]{@{}c@{}} $\bm{E_\mathrm{hull}}$ \\ (eV/atom)\end{tabular}} &
    \textbf{\begin{tabular}[c]{@{}c@{}} $\bm{E_\mathrm{G}}$ \\ (eV)\end{tabular}} &
    \textbf{\begin{tabular}[c]{@{}c@{}} $\bm{E_\mathrm{G}^\mathrm{d}}$ \\ (eV)\end{tabular}} &
    \textbf{\begin{tabular}[c]{@{}c@{}} $\bm{m^*_\mathrm{e}}$\end{tabular}} &
    \textbf{\begin{tabular}[c]{@{}c@{}} $\bm{m^*_\mathrm{h}}$\end{tabular}} \\ \midrule

WS & $P3m1$ & 4 & 0.0 & 2.47 & 3.10 & 7.30 & 1.69 \\
LC & $Pca2_1$ & 16 & 0.0106 & 1.63 & 1.63 & 1.33 & 1.87 \\
d-HX & $P2_1/c$ & 16 & 0.0294 & 2.62 & 2.71 & 3.88 & 2.16 \\
d-WS & $Cm$ & 32 & 0.0312 & 2.18 & 2.18 & 1.56 & 1.25 \\
BN & $Cm$ & 16 & 0.0327 & 2.01 & 2.01 & 1.36 & 1.49 \\
d-RS & $Pc$ & 16 & 0.0373 & 2.22 & 2.47 & 3.41 & 2.17 \\
WZ & $Pmc2_1$ & 8 & 0.0496 & 2.53 & 3.23 & 0.62 & 3.62 \\
RS & $I4_1/amd$ & 16 & 0.0807 & 1.15 & 1.87 & 0.83 & 1.96 \\
ZB & $P\bar{4}m2$ & 4 & 0.0857 & 2.04 & 3.03 & 0.52 & 1.59 \\ \bottomrule
\end{tabular}}


\footnotesize{\textsuperscript{†} Number of atoms in primitive unit cell 
}
\end{table}

\subsection{Identification of possible \ce{ZnZrN2} polymorphs}

Although only the WS phase is reported in the Materials Project database (\ce{ZrZnN2}, "mp-1014244"),\cite{osti_1337299, jain2013commentary} this unexplored Zn-Zr-N phase space could in principle host a variety of different structures. Many methods exist to determine possible polymorphs and predict synthesizable compounds,\cite{woodley2008crystal} ranging from simple ionic substitution,\cite{hautier2010data} to kinetically limited minimization (KLM)\cite{sharan2021computational}, \textit{ab initio} random structure searches (AIRSS),\cite{pickard2011ab} or more expensive evolutionary and genetic algorithms.\cite{oganov2011evolutionary} Since no single method is fully representative of configurational space, we use the combined methods of KLM,\cite{sharan2021computational} random structure searching,\cite{stevanovic2016sampling} and ionic substitution \cite{greenaway2020combinatorial} to predict 28 unique possible ordered \ce{ZnZrN2} polymorphs (three of which have been added to the NRELMatDB \cite{stevanovic2012correcting, lany2013bandstructure, lany2015semiconducting}). Most of these polymorphs have unit cells of 16 atoms or fewer and represent various orderings, and thus are referred to as "ordered" polymorphs herein.

The resulting 28 polymorphs are classified into six distinct structure prototype classes, with representative crystals for each of these structure prototypes depicted in \autoref{fig:prototypes}(a), and adopt a naming convention from binary analogs as follows: rocksalt-derived ("RS") is an fcc anion sublattice with cations in O\textsubscript{h}-coordinated voids, wurtzite-derived ("WZ") exhibits a structurally face-centered tetragonal anion lattice with tetrahedral coordinated cations, zincblende-derived ("ZB", i.e. chalcopyrite) is an fcc anion sublattice with cations in every other tetrahedral void, wurtsalt ("WS") presents alternating layers of octahedrally coordinated Zr and tetrahedrally coordinated Zn (as discussed previously), and boron-nitride-derived ("BN") exhibits hexagonal sheets of various stackings (similar to graphite, but a 3D structure with \textit{M}-N bonds between c-axis layers and the key distinction that the c-axis bonds lengths are nearly equal to the in-plane bond lengths\cite{limpijumnong2001theoretical}). An additional compound, with alternating 2D layers corresponding to layers of the mineral litharge (PbO, with a space group $P4/nmm$) and the mineral massicot (PbO, with a space group $P2_1ca$), respectively, we name with the amalgam ``lithicot'' ("LC"; we were also unable to locate an existing mineral name). The heavily distorted versions of three of these prototypes are categorized separately, with the prefix ``d-'', using a tolerance developed from a structure matching algorithm; see Supplemental Materials (SM) for classification scheme details. The structure class "d-HX" (HX = hexagonal) represents structures that are distortions between BN and WZ, which are related to one another through a displacive transformation. \ce{Zn3N2} crystallizes in an anti-bixbyite-derived phase (``BX''), which is observed experimentally, but this structure is not included in our set of prototypes since deriving an analogous topotactic \ce{ZnZrN2} BX structure requires removing atoms and is not trivial.

The resulting 0 K formation energy of a given ordered polymorph in structure class, $k$, is referred to as $\Delta H_k^\mathrm{ord}$ (e.g. $\Delta H_\mathrm{WS}^\mathrm{ord}$, $\Delta H_\mathrm{RS}^\mathrm{ord}$, etc.). \autoref{fig:prototypes}(b) plots the resulting energy ordering of the 28 ordered structures, with polymorphs grouped by structure type and ``$E_\mathrm{hull}$'' indicating the energy above the convex hull, i.e. the difference between the computed $\Delta H_k^\mathrm{ord}$ and the ground state hull. The SCAN functional confirms a predicted ground state WS ($P3m1$) that lies on the convex hull, corroborating the literature.\cite{hinuma2016discovery, sun2016thermodynamic} Other WS polymorphs ($P6_3mc$, $P\bar{3}m1$) are low in energy, ranging from 0 to 0.025 eV/atom, and the LC structures ($Pca2_1$, $Iba2$) are the next-lowest in energy. RS polymorphs are the highest in energy, with $E_\mathrm{hull}$ values ranging from 0.080 to 0.156 eV/atom. PBE and PBE+U yield similar energy orderings (see SM), although LC is the predicted ground state for PBE without a Hubbard U correction (see SM).

The $E_\mathrm{hull}$ values of the lowest energy ordered structure in each prototype class, as well as their calculated band gaps ($E_{\mathrm{G}}$) and electron and hole effective masses ($m^*_{\mathrm{e}}$ and $m^*_{\mathrm{h}}$) from SCAN, are reported in \textbf{\autoref{tab:polymorph_energies}}, with the full list in the SM (note that reported $E_{\mathrm{G}}$ are Kohn-Sham gaps calculated with SCAN, which systematically underestimates the true band gap \cite{borlido2019large}). Optoelectronic properties vary significantly by structure. Most polymorphs have indirect gaps except for the LC structures, most of the BN, some distorted structures, and one RS. The WZ $Pna2_1$ polymorph exhibits the largest band gap ($E_{\mathrm{G}}$ $\approx$ 2.99 eV with SCAN, see SM), followed by d-HX, WS, ZB, d-RS and d-WS with $E_{\mathrm{G}} >$2 eV, while RS has among the lowest band gaps ($\sim$0--1.67 eV, depending on cation ordering). Notably, the lowest-energy WS $P3m1$ polymorph has an exceptionally low $m^*_{\mathrm{h}}$ ($<$2) compared to  $m^*_{\mathrm{e}}$ ($<$7) while retaining a wide direct band gap, $E_{\mathrm{G}}^\mathrm{d} > $3 eV. This combination of electronic structure properties is unique among all the considered polymorphs, and is rare for other chemistries outside of the \ce{ZnZrN2} material system.

\subsection{Synthesis of metastable phases}

\begin{figure}[!ht]
    \centering
    \includegraphics[width=80mm]{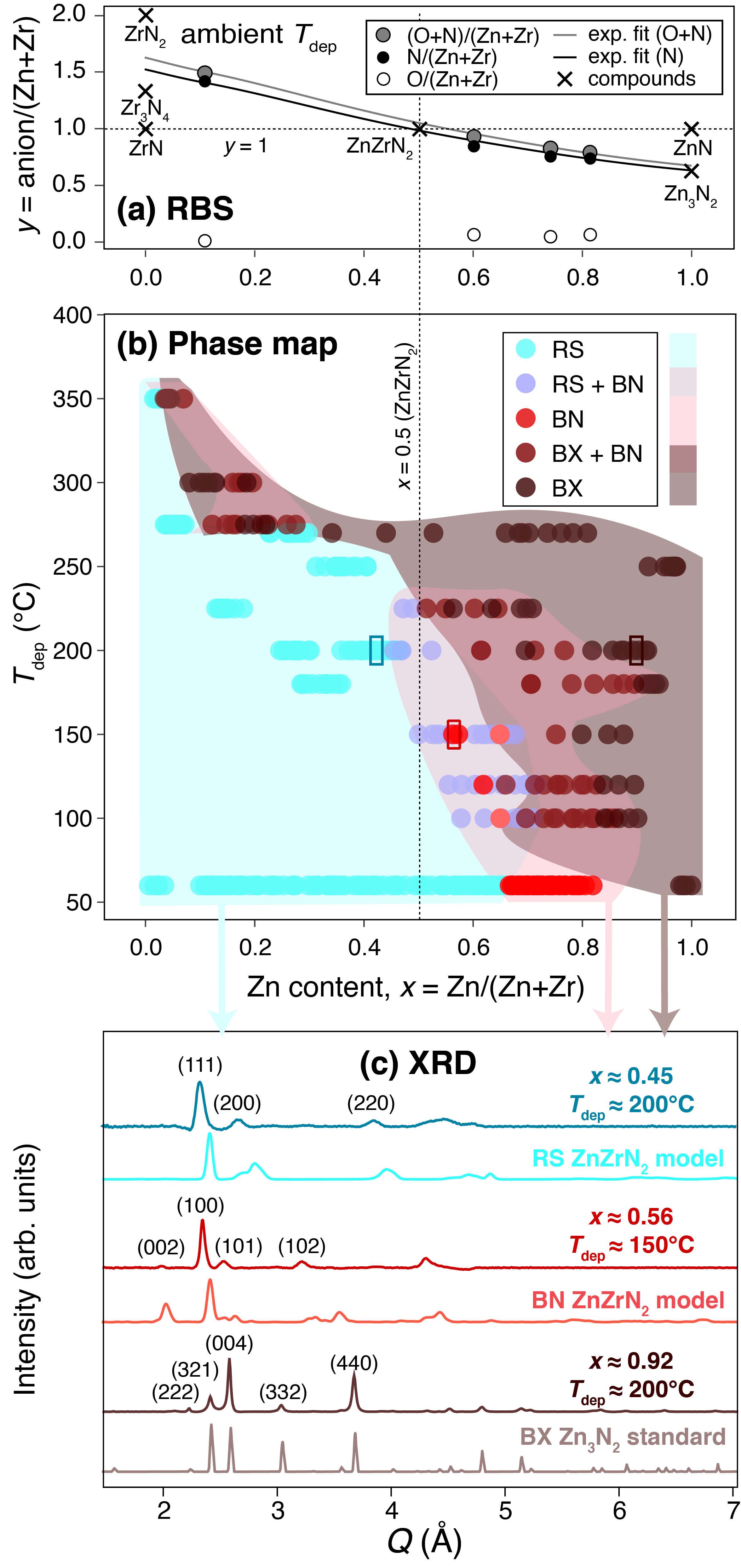} 
    \caption{(a) RBS measurements of anion composition as a function of cation composition, with exponential fits polotted. (b) Map of over 400 samples in Zn-Zr-N experimental phase space, with cation composition $x=$ Zn/(Zn+Zr) on the x-axis and calibrated deposition temperature $T_\mathrm{dep}$ on the y-axis. (c) Representative XRD patterns for 3 samples marked with rectangles in (b), with "modeled" \ce{ZnZrN2} RS and BN from ensemble averages of XRD patterns from the polymorph sampler method for \ce{ZnZrN2} compositions, and "standard" \ce{Zn3N2} anti-bixbyite (BX) from the Materials Project database plotted with a smearing bandwidth.}
    \label{fig:phase_map}
\end{figure}

Despite the existence of at least 19 predicted polymorphs with lower 0 K DFT formation energies, an RS phase with a high $E_\mathrm{hull}$ is experimentally synthesized at low deposition temperatures ($T_\mathrm{dep}$) and \ce{ZnZrN2} stoichiometry. Using combinatorial sputter synthesis,\cite{green2013applications} a set of approximately 400 samples in the \ce{Zn_$x$Zr_{1-$x$}N_$y$} ternary alloy system is grown, with cation concentration ranging from 0 $\leq x \leq$ 1 and growth temperature $T_\mathrm{dep}$ ranging from ambient to 500\degree C. \textbf{\autoref{fig:phase_map}}(a) depicts RBS anion-to-cation ratio, $y =$ anion/(Zn+Zr) with anion = (O+N), N, or O, as a function of cation ratio, $x =$ Zn/(Zn+Zr), for a set of representative samples grown at ambient temperature. RBS corroborates the cation concentration measured by XRF and indicates N-rich compositions in Zn-poor samples, N-poor compositions in Zn-rich samples, and approximately stoichiometric N at the \ce{ZnZrN2} composition of interest. Additionally, a small but nonzero presence of O is detected, likely substituting for N and plausibly residing on the film surface ($\sim$0.3 at. \% in Zn-poor samples, $\sim$5 at. \% in Zn-rich samples due to reaction of zinc nitride with ambient atmosphere; see SM). An exponential fit suggests our samples have crystallized near the \ce{Zr3N4}--\ce{Zn3N2} tieline, as indicated by the exponential fit to RBS referenced to the crossed markers, with an approximate stoichiometry of \ce{Zn_$x$Zr_{1-$x$}N_{$y$}} where $y \approx (4-2x)/3$. This system could alternately be expressed as "\ce{Zn_{1+$x$}Zr_{1-$x$}N_{2+$y$}}" to emphasize off-stoichiometry from \ce{ZnZrN2} (see SM). For simplicity and generality we will refer to experimental alloys as "\ce{Zn_$x$Zr_{1-$x$}N_{$y$}}" herein since multiple experimental phases are observed, and focus on varying $x$ since the anion composition $y$ is not intentionally tuned. 

Using high-throughput synchrotron XRD, and referencing the XRD patterns from the set of predicted polymorphs, the structural phase space is mapped in \autoref{fig:phase_map}(b) by linking the structure of each of the $\sim$400 samples to its corresponding ternary composition and calibrated deposition temperature, $T_\mathrm{dep}$. Rather than crystallizing in its predicted ground state WS structure at and around the \ce{ZnZrN2} ($x$ = 0.5) composition, synchrotron XRD demonstrates predominantly an RS crystal structure, corresponding to an $E_\mathrm{hull}$ of at least 0.08 eV/atom according to \autoref{fig:prototypes}. The transformation to a second phase is observed at higher $x$ compositions, a hexagonal structure corresponding to BN (or possibly d-HX), and a third anti-bixbyite-derived phase (``BX'') is observed at the highest $x$ concentrations near \ce{Zn3N2} (simulating BX \ce{ZnZrN2} is nontrivial and is not performed here). At higher $T_\mathrm{dep}$, there are regions of mixed phases of these three polymorphs, perhaps due to miscibility or Zn volatility. We attempted to synthesize Zn-rich samples at $T_\mathrm{dep}$ > 250\degree C, but no such samples were realized due to the high vapor pressure of Zn under our growth conditions (see SM).

Figures \ref{fig:phase_map}(b) and (c) represent the key structural observations in \ce{Zn_$x$Zr_{1-$x$}N_$y$}. First, an RS-derived phase dominates from $x$ = 0 up to a threshold $x$ value, which is approximately $x \approx 0.66$ at ambient temperature growth conditions ($T_\mathrm{dep} \approx$ 65\degree C, bottom of figure) and which drops as $T_\mathrm{dep}$ increases. An XRD pattern for a representative RS sample of $x\approx$ 0.45 is depicted in (c) in dark teal, compared to a modeled RS XRD pattern in light teal, simulated as ensemble-averages from the polymorph sampler. The RS-derived phase at $x$ = 0 is more N-rich than RS ZrN, so we refer to it as \ce{ZrN_$y$} ($y$ > 1). This could in principle be a single or mixed phase of RS \ce{ZrN_$y$} ($y$ > 1), \ce{Zr3N4},\cite{guo2010first, klumdoung2012variation} \ce{ZrN2}, \ce{Zr3N2},\cite{yu2017first, sun2017thermodynamic} or \ce{Zr2N3-$y$} ($y$ = 0.34),\cite{clarke1999structure} with possible defect-mediated or oxygen-induced stabilization; in-depth investigation of this phase is beyond our scope. As the Zn content increases, the XRD peaks around $Q$ = 2.3 and 2.65 \AA, which correspond to RS (111) and (200), respectively (indices from the ZrN RS structure; see SM), shift to higher $Q$ values, with the former strengthening and the latter weakening. This trade-off is likely due to shifts in texturing, as also commonly observed in other ternary nitrides.\cite{bauers2019composition}

At a threshold composition ($x \approx$ 0.66 at ambient temperature), there is a phase transformation to a hexagonal BN-derived structure. \autoref{fig:phase_map}(c) depicts a representative BN diffraction pattern for a sample with $x \approx$ 0.56 in dark red, with diffraction peaks at $Q$ values of $\sim$2.10, 2.45, and 2.55 \AA \space corresponding to BN (002), (100), and (101) reflections, respectively. This transformation occurs at lower $x$ values for samples grown in the approximate range 100\degree C $\lesssim T_\mathrm{dep} \lesssim$ 225\degree C, with a large region of mixed phase RS and BN ("RS + BN").

At a second threshold composition ($x \gtrsim$ 0.8 at ambient temperature, and lower $x$ for high $T_\mathrm{dep}$), a second phase transition occurs to the BX phase that phase holds until $x$ = 1 with a stoichiometry of approximately \ce{Zn3N2}. The presence of BX \ce{Zn3N2} corroborates literature reports,\cite{partin1997crystal} and may be enabled by Zr\textsubscript{Zn} antisite stabilization across phase space. There are several regions of phase-segregated BX as well, in particular at $T_\mathrm{dep} >$ 250 \degree C. Here, films are likely completely phase-separating into binaries of RS \ce{ZrN_$y$} and BX \ce{Zn3N2}, though it is also plausible that a BX-derived phase of \ce{ZrN_$y$} or Zr-rich Zn-Zr-N has formed and is responsible for the BX reflections.

In summary, RS and BN are synthesized near the \ce{ZnZrN2} composition ($x$ = 0.5) and BX at high $x$, but no WS phase is observed. The measured and simulated XRD patterns correspond very well, except for offsets in $Q$ that are a consequence of errors in DFT lattice constants or experimental artifacts (e.g. off-stoichiometry, possible residual strain, sample misalignment). It is notable that in this alloy system \ce{Zn_{$x$}Zr_{1-$x$}N_{$y$}}, the presence of a lower-density hexagonal phase (BN, here) located between two higher density cubic phases (RS and BX, here) is indicative of a phenomenon in heterovalent heterostructural alloys called "negative pressure" polymorphs,\cite{siol2018negative, woods2019combinatorial} and this space warrants further exploration.


\subsection{Statistical sampling of thermodynamically accessible polymorphs}

The synthesis of metastable polymorphs (RS and BN) rather than the predicted ground state (WS) is not particularly surprising; DFT is a 0 K, thermodynamic equilibrium modeling approach of bulk systems while sputtering is a high effective-temperature, non-equilibrium synthesis approach of thin films, and so the two methods are not necessarily compatible. Despite these incompatibilities, DFT often \textit{does} correctly predict sputtered crystal structures, for example in other ternary nitrides\cite{sun2019map} or in numerous oxide compounds, and thus is commonly used for simulating such materials. However, there are also other cases in the literature where the predicted DFT ground state is not synthesizable via sputtering or where sputtering can access metastable states. For example, ternary nitride \ce{ZnMoN2} is predicted in a layered structure but synthesized in a WZ structure,\cite{arca2018redox} \ce{Mg2NbN3} is predicted in a layered structure but synthesized in a RS structure,\cite{bauers2019ternary} and \ce{Zn2SbN3} and \ce{Mg2SbN3} are metastable with respect to decomposition into \ce{N2} yet both can be made by sputtering.\cite{arca2019zn, heinselman2019thin}

It is still not understood, for a given system, whether the DFT ground state will ultimately be synthesizable as a sputtered thin film or whether a higher-energy polymorph will crystallize instead, and in each case why or why not. Modeling sputtering from first principles is computationally difficult (e.g. time-dependent or Monte Carlo simulations), and is further complicated since sputtered films tend to decompose before equilibrium is reached. The computational analysis herein seeks to contextualize our experimental findings by approximating whether metastable states could be accessible experimentally using non-equilibrium synthesis techniques such as sputtering. These computational methods and the following discussion are \textit{not} aimed to show that WS cannot be synthesized --- it may very well be possible to synthesize WS under different conditions --- but rather, we provide a rationale for why metastable phases have been stabilized under these experimental conditions.

\begin{figure}[!ht]
    \centering
    \includegraphics[width=65mm]{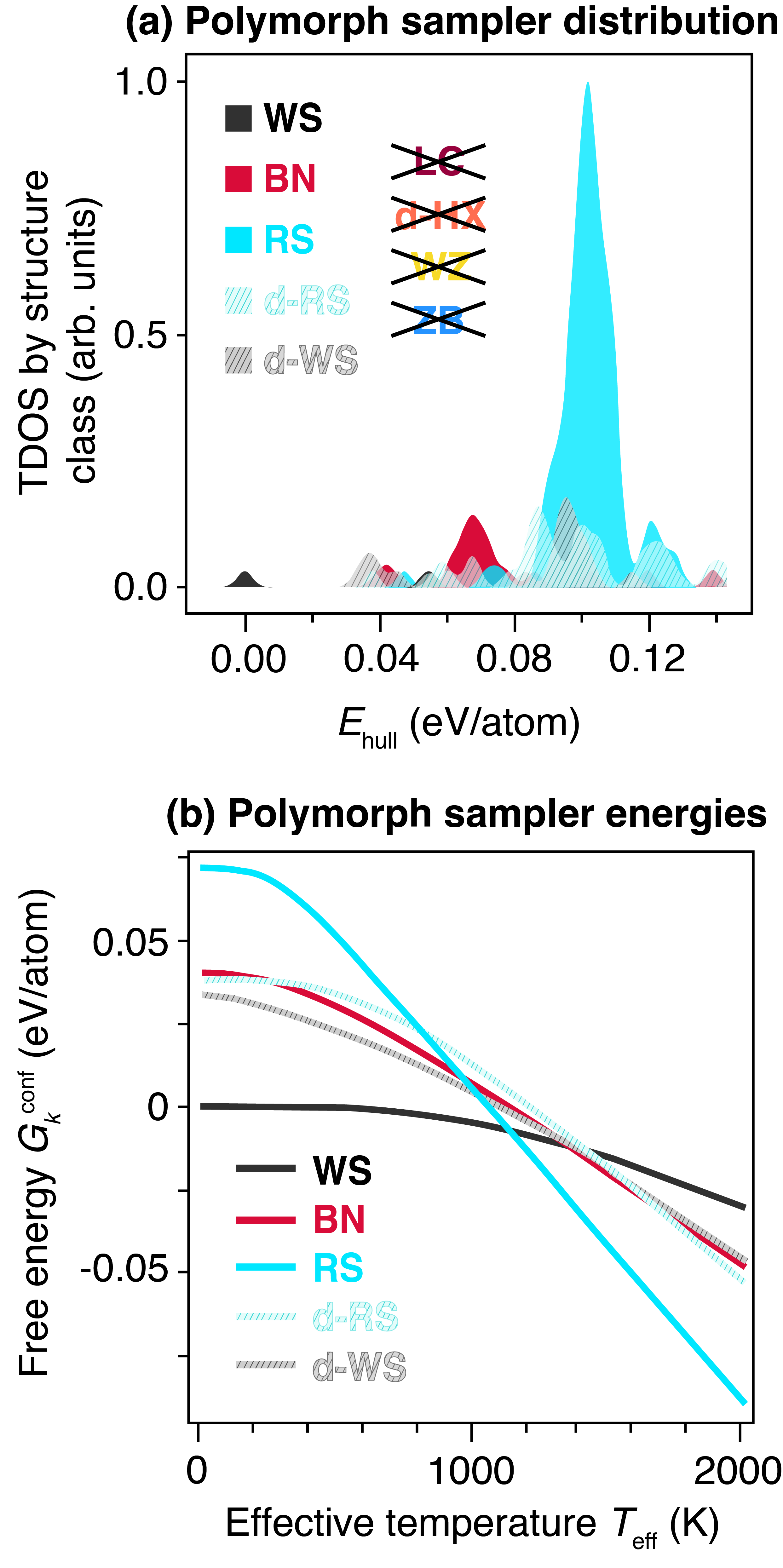}
    \caption{(a) Thermodynamic density of states (TDOS) distribution of sampled \ce{ZnZrN2} polymorphs from random structure searching, demonstrating a dominant RS peak, as calculated from \autoref{eq:prob}. Crossed-out classes depicting the absence of several predicted polymorph phases. (b) Free energy $G_k^\mathrm{conf}$ as a function of effective temperature $T_\mathrm{eff}$, shifted such that the ground state at $T_\mathrm{eff}$ = 0 K is at the origin, calculated from the TDOS using \autoref{eq:free-energies}.}
    \label{fig:free_energy}
\end{figure}

It has previously been shown that treating a spectrum of structures generated by random structure sampling\cite{pickard2011ab} as a proper statistical ensemble can determine the experimental realizability of metastable crystalline polymorphs phases (e.g. MgO, ZnO, \ce{SnO2}, and Si)\cite{stevanovic2016sampling, jones2017polymorphism} as well as the structural features of glasses.\cite{jones2020glassy} Here, 5,000 \ce{ZnZrN2} RSL structures are generated with random structure sampling, and the resulting spectrum of polymorph structures is split into prototype classes with the same underlying space group (see SM). Next, the ensemble probability $P_k$ of every individual class is evaluated as:

\begin{equation}
    \label{eq:prob}
    P_k \approx \frac{\sum_{n=m}^{m+n_k}\omega_n \upe^{-\frac{E_{\mathrm{hull,}n}}{k_\mathrm{B} T_\mathrm{eff}}}}{\Xi} = \frac{\Xi_k}{\Xi}
\end{equation}

\noindent where $k$ represents different prototype classes, $n$ counts polymorph structures within a given class, $\omega_n$ is the frequency of occurrence of a structure $n$ belonging to class $k$, $E_{\mathrm{hull,}n}$ is the formation energy per atom relative to the ground state, and $T_\mathrm{eff}$ is the effective temperature (see \autoref{eq:free-energies} below). $\Xi_k$ and $\Xi$ are the partial and the full partition functions, respectively. The former pertains only to the $k$-class of structures, while the latter is evaluated for all RSL structures. We use the ``$\approx$'' symbol to emphasize that the ensemble probabilities from \autoref{eq:prob} are the approximate versions of the true ensemble probabilities (only configurational degrees of freedom are sampled). All of the approximations adopted in the polymorph sampler approach are discussed and analyzed at length by Jones and Stevanoviç.\cite{jones2020glassy}

The thermodynamic density of states (TDOS; i.e., number of structures per energy unit) resulting from the random structure sampling, normalized and resolved by structure class, is shown in \textbf{\autoref{fig:free_energy}}(a). Two features become immediately apparent. First, consistent with \autoref{fig:prototypes}(b) and literature reports, the ground state WS structure is correctly identified (dark grey), but its corresponding frequency of occurrence and associated TDOS are very small. Second, RS structures have the largest TDOS peak (cyan), concentrated in the narrow window of 0.08--0.12 eV/atom. This suggests the flexibility of RS to accommodate cation disorder in a relatively narrow energy interval. Additional classes of disordered structures with more significant occurrence are the BN (red) and the two highly distorted d-RS and d-WS classes (striped), with the latter also including structures with more than one tetrahedral layer sandwiched between the octahedrally-coordinated layers of the WS structures. It is also important to note that none of the other previously discussed structure classes appear in the RSL structures, including the LC and WZ classes with DFT formation energies lower than that of RS. This indicates that all possible structures in these absent classes exhibit very "narrow" local minima in configurational energy space, leading to a very low probability of occurrence. The same is true for a relatively large number of very low symmetry structures (space groups $P1$ and $P\bar{1}$) typically obtained in random structure sampling, but since none of these structures occur in large numbers they become statistically insignificant compared to those depicted in \autoref{fig:free_energy}.

The TDOS from \autoref{fig:free_energy}(a) allows evaluating ensemble probabilities per \autoref{eq:prob} and associated partial partition functions $\Xi_k$. These are used to evaluate the “configurational” free energies $G_k^\mathrm{conf}$ of the corresponding structure types using the standard statistical mechanics equation:

\begin{equation}
    \label{eq:free-energies}
    G_k^\mathrm{conf} (T_\mathrm{eff}) = -k_\mathrm{B} T_\mathrm{eff} \ln \Xi_k
\end{equation}

\noindent where $k_\mathrm{B}$ is the Boltzmann constant.

$T_\mathrm{eff}$ is the "effective temperature," defined in the literature as the thermodynamic temperature where a material grown \textit{in} equilibrium would have the same degree of disorder as the same material grown \textit{out} of equilibrium (e.g. by sputtering).\cite{ndione2014control, cordell2021probing} Effective temperature $T_\mathrm{eff}$ can be thought of as a proxy for disorder, such that higher $T_\mathrm{eff}$ represents higher disorder in a given material. The $T_\mathrm{eff}$ models configurational disorder as typically seen in non-equilibrium synthesis. $T_\mathrm{eff}$ and $T_\mathrm{dep}$ are not directly comparable; rather, low deposition temperatures generally correspond to high $T_\mathrm{eff}$, because kinetic limitations inhibit enthalpy-driven ordering (see "Tolerance to off-stoichiometry informs phase transitions"). Accordingly, the corresponding free energy $G_k^\mathrm{conf}$ excludes non-configurational free energy contributions such as vibrational contribution (see next section, "Vibrational contributions are negligible").\cite{cordell2021probing} Also, the ideal gas free energy of \ce{N2}, which is otherwise by far the largest finite-temperature free energy contribution under thermodynamic equilibrium conditions (up to several eV, depending on temperature and partial pressure), does not apply in sputtering synthesis, where high non-equilibrium nitrogen chemical potentials up to $\Delta \mu_\mathrm{N}$ = +1.0 eV can be achieved.\cite{caskey2014thin}

The resulting $T_\mathrm{eff}$ dependence of the free energy $G_k^\mathrm{conf}$, displayed in \autoref{fig:free_energy}(b), clearly shows that at low $T_\mathrm{eff}$ the lowest free energy structure is the ground state WS structure, consistent with \autoref{fig:prototypes}. However, at $T_\mathrm{eff} \gtrsim$ 1150 K, the disordered RS becomes the most favorable structure due to the large gain in configurational entropy.\cite{rost2015entropy} In the temperature range 1300--1600 K, the WS structure gives way to disordered BN as the second most favorable structure, while at still higher temperatures the d-RS becomes the most favorable. This structure, if mixed with RS, would be experimentally difficult to distinguish from RS using XRD because of their very similar diffraction patterns.

In summary, at higher effective temperatures the polymorph sampler ensemble treatment suggests the following ordering of structures according to $G_k^\mathrm{conf}$, from lowest to highest: (1) RS, (2) distorted RS (d-RS), (3) BN, (4) distorted WS (d-WS), and (5) WS. This is consistent with our experimentally observed XRD patterns that are compared with the ensemble-averaged patterns in \autoref{fig:phase_map}(c). We reiterate that $T_\mathrm{eff}$ is representative of \textit{effective} temperature rather than $T_\mathrm{dep}$, the substrate temperature during sputter synthesis. Previous studies have suggested that sputter deposition occurs at $T_\mathrm{eff}$ higher than 1150 K in ternary nitrides; in fact, $T_\mathrm{dep}$ has been shown to scale \textit{inversely} with $T_\mathrm{eff}$ for sputtered films (where $T_\mathrm{dep} \lesssim$ 600\degree C) since strong kinetic limitations at low $T_\mathrm{dep}$ induce a high degree of disorder (see SM).\cite{fioretti2018exciton, lany2017monte} Therefore, since computed phases at high $T_\mathrm{eff}$ correspond to phases grown at low $T_\mathrm{dep}$ in \autoref{fig:phase_map}, these ensemble probabilities and free energies support the observed behavior in sputter-deposited samples: the RS phase is stabilized and the WS phase is destabilized.

%



\begin{figure}[!ht]
    \centering
    \includegraphics[width=90 mm]{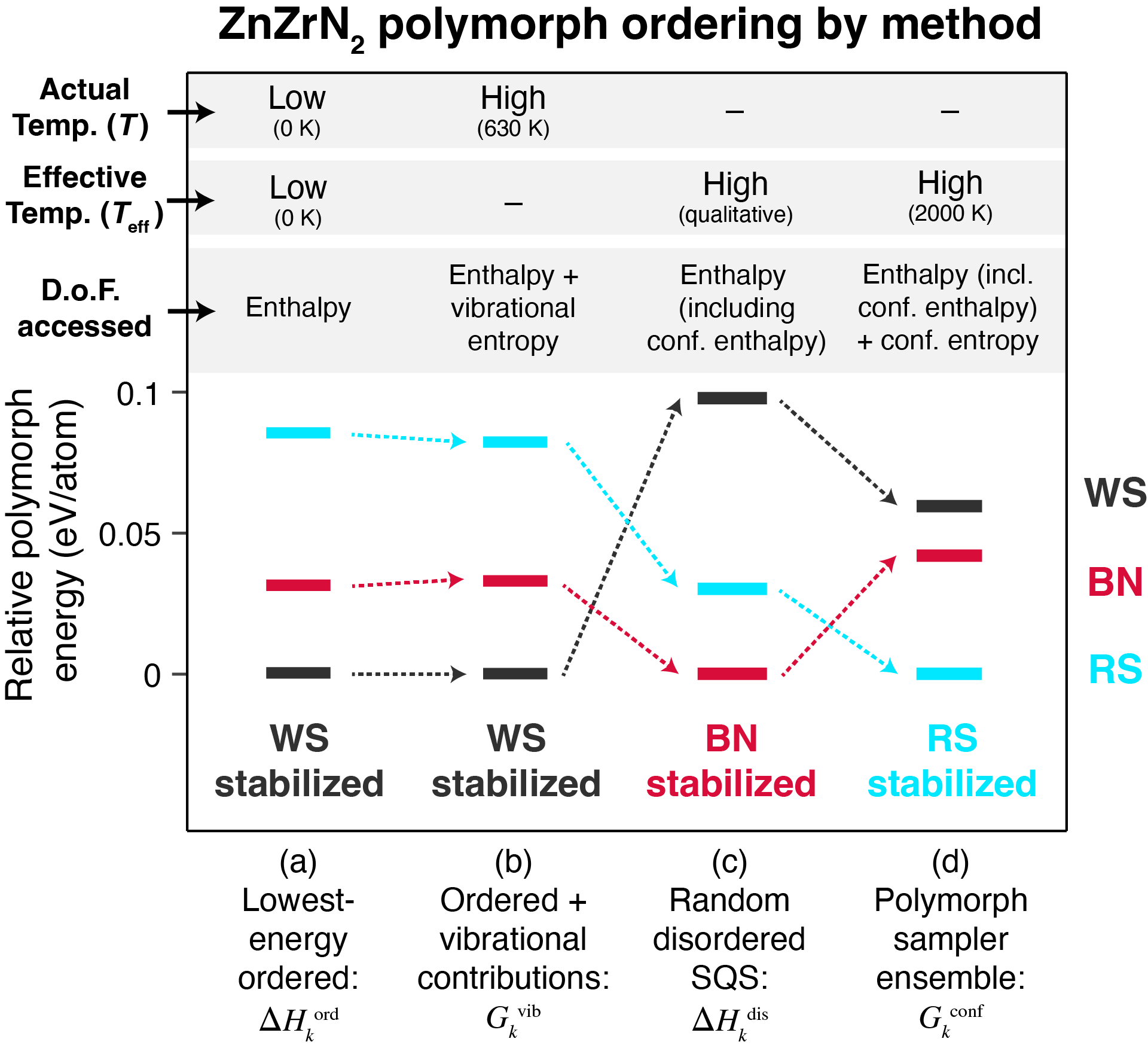}
    \caption{Comparison of relative polymorph energy orderings by calculation method for WS, BN, and RS structure classes, from left to right: (a) enthalpies of lowest energy ordered structures $\Delta H_{k}^\mathrm{ord}$ (i.e. $E_\mathrm{hull}$ in \autoref{fig:prototypes}), (b) free energies including vibrational contributions of ordered structures $G_k^\mathrm{vib}$ at $T$ = 630 K, (c) enthalpies of disordered SQS structures $\Delta H_{k}^\mathrm{dis}$ (from \autoref{fig:sqs}), and (d) the polymorph sampler free energies $G_k^\mathrm{conf}$ at $T_\mathrm{eff}$ = 2000 K (from \autoref{fig:free_energy}). A qualitative metric ("low" or "high") is reported for actual temperature $T$ and effective temperature $T_\mathrm{eff}$ (a proxy for cation disorder, as explained in the text), and accessed degrees of freedom (D.o.F.) are labeled ("conf" corresponds to configurational D.o.F.). $T$ = 630 K is selected for $G_k^\mathrm{vib}$ to represent the highest deposition temperature ($T_\mathrm{dep}$) probed experimentally in this paper, rounded up to the nearest 10 K. Energy is referenced on the y-axis with respect to the lowest formation energy for a given method, and arrows are a guide to the eye.}
    \label{fig:comparison}
\end{figure}

\subsection{Vibrational contributions are negligible}

We have highlighted the role of configurational degrees of freedom in this system, but it is also important to assess the magnitude of other energetic contributions, in particular vibrational contributions, to assess whether they significantly change energy ordering. Here, we use DFPT to estimate energetic contributions from vibrational degrees of freedom for the lowest energy RS, BN, and WS phases of \ce{ZnZrN2}, and report the resulting Gibbs free energy $G_k^\mathrm{vib}$ (details provided in the SM). It is noted that the $G_k^\mathrm{vib}$ is a function of the actual synthesis temperature $T$ (i.e., $T_\mathrm{dep}$ up to about $\sim$630 K here), rather than the effective temperature $T_\mathrm{eff}$ discussed in the polymorph sampler approach.\cite{togo2015phonopy, cordell2021probing} These results show that RS is somewhat destabilized with respect to BN at very high temperatures ($T\approx$ 1800 K), but across all assessed temperatures WS is still the lowest energy structure compared to RS or BN. At the highest experimentally probed temperature ($T_\mathrm{dep}$ = 350\degree C, i.e. $\sim$630 K), the relative change in the RS and BN polymorph energy due to vibrational effects is approximately 3--4 meV/atom, which is much smaller than the polymorph sampler energy differences observed at high $T_\mathrm{eff}$ in \autoref{fig:free_energy}(b). These energy differences are depicted in \textbf{\autoref{fig:comparison}} by comparing the relative energy ordering between (a) $\Delta H_k^\mathrm{ord}$, (b) $G_k^\mathrm{vib}$, and (d) $G_k^\mathrm{conf}$, and are elaborated upon in the discussion section. Therefore, vibrational effects to not explain the stabilization of BN and RS over the WS phase observed in our experiments. Rather, by comparing to the configurational contributions to free energy, we show RS and BN are stabilized and WS is destabilized at high temperature by \textit{configurational} degrees of freedom rather than vibrational degrees of freedom.

\section{Discussion and implications}

We have synthesized the RS-derived and BN-derived \ce{ZnZrN2} structures rather than WS, which is the DFT-predicted ground state at 0 K, and have used a statistical polymorph sampler to explain these results by demonstrating that RS becomes the lowest energy polymorph at high effective temperatures. However, this does not explain the physical principle behind why certain structures are stabilized or destabilized upon disorder, nor why BN is synthesized at Zn-rich compositions. Inspection of structures in \autoref{fig:prototypes}(a) indicates that the ordered polymorph structures with lowest $\Delta H_k^\mathrm{ord}$ exhibit unique, inequivalent cation coordination environments for Zr and Zn (WS, LC), while higher formation energy structures have similar, equivalent coordination environments for all cation sites (RS, BN, WZ, ZB). This is demonstrated in \textbf{\autoref{fig:sqs}}(a) with partially occupied WS and RS structures.

Our hypothesis is that cation disordering during synthesis---enabled by rapid condensation from the vapor state to the solid state in physical vapor deposition (PVD) techniques such as sputtering---favors structures with similar cation coordination environments, thus lowering the probability of the formation of WS. To examine this hypothesis and explore how the polymorph sampler results pertain to other systems, we pursue two high-throughput computational approaches. First, we develop a descriptor to interpret the results of the polymorph sampler within the framework of random cation disorder, and second, we estimate formation energies of ordered \ce{Zn_{$x$}Zr_{1-$x$}N_{$y$}} with varied cation ratios $x$ to assess the effects of off-stoichiometry in relation to experimental phase space.

\subsection{Tolerance to disorder influences synthesizability at high effective temperatures}

In practice, cation disorder is ubiquitous in ternary nitrides,\cite{quayle2015charge, lany2017monte, schnepf2020utilizing} especially in materials synthesized at high effective temperatures (as present in sputtering). Thus, modeling small, cation-ordered unit cells as in \autoref{fig:free_energy} may not adequately capture energetic information in these systems. Here, the energetic effects of random cation disorder in \ce{ZnZrN2} polymorph structures are explicitly considered by generating random disordered structures in each structure class using the SQS method, as described previously. For each structure class $k$, these resulting formation energies all are within $\sim$0.010 eV/atom of one another; these energies are then ensemble-averaged to best represent the formation energy of a randomly disordered phase, referred to as $\Delta H_{k}^\mathrm{dis}$.

\begin{figure}[!ht]
    \centering
    \includegraphics[width=90 mm]{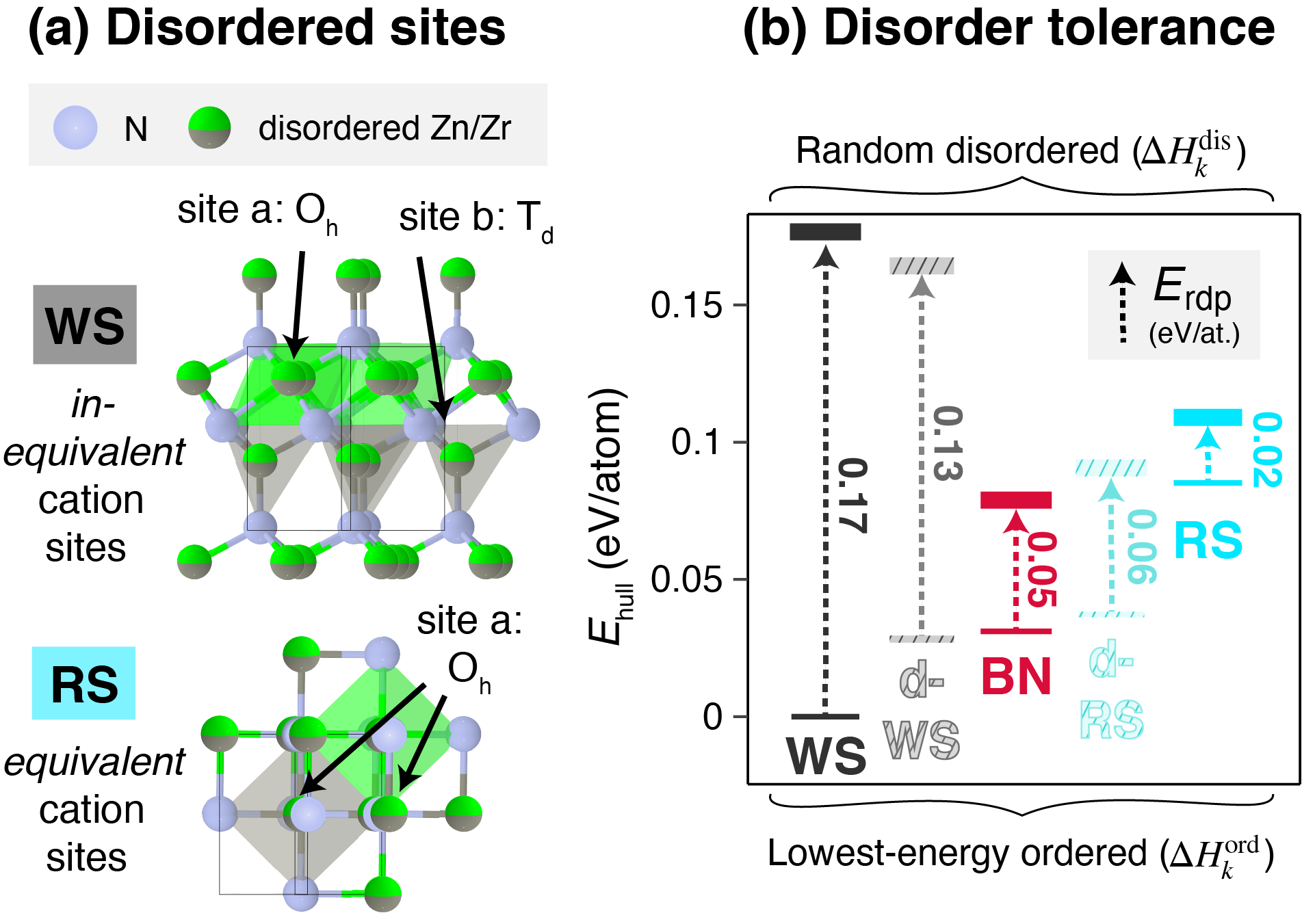}
    \caption{(a) Schematic of the WS and RS derived structures with cation site disorder as an example of structures with inequivalent and equivalent cation sites respectively. (b) Comparison of formation energies of the lowest energy ordered polymorph structure ($\Delta H_{k}^\mathrm{ord}$) and of the random disordered structure ($\Delta H_{k}^\mathrm{dis}$) for each prototype class plotted in \autoref{fig:free_energy}. The $E_\mathrm{rdp}$ descriptor is shown, with a lower $E_\mathrm{rdp}$ correlating to the higher disorder tolerances in the BN and RS structures that are observed experimentally (see SM).}
    \label{fig:sqs}
\end{figure}

\autoref{fig:sqs}(b) compares $\Delta H_{k}^\mathrm{dis}$ to the $\Delta H_{k}^\mathrm{ord}$ of the lowest energy ordered structures for the five structure classes that emerge from the polymorph sampler (see SM for others), referenced to $\Delta H_\mathrm{WS}^\mathrm{ord}$ and reported as $E_\mathrm{hull}$. The random disordered WS and d-WS structures have high formation energies, with $\Delta H_\mathrm{WS}^\mathrm{dis}$ over 0.17 eV/atom. In contrast, $\Delta H_\mathrm{BN}^\mathrm{dis}$ is lowest of all disordered structures. Although RS does not have the lowest $\Delta H_k^\mathrm{dis}$, $\Delta H_\mathrm{RS}^\mathrm{dis}$ is very similar (within 0.025 eV/atom) to $\Delta H_\mathrm{RS}^\mathrm{ord}$, and thus has a lower energetic penalty to disorder. To assess this "disorder tolerance" for a given structure class $k$, we introduce a new descriptor, the "random disordered polymorph energy"  $E_\mathrm{rdp}$:

\begin{equation}
    \label{eq:Erdp}
    E_\mathrm{rdp}(k) = \Delta H_{k}^\mathrm{dis} - \Delta H_{k}^\mathrm{ord} .
\end{equation}

\noindent RS and BN, the structures that have been experimentally synthesized, have the lowest $E_\mathrm{rdp}$ values. Physically, since the 0 K DFT formation energy is an approximation of formation \textit{enthalpy}, the $E_\mathrm{rdp}$ represents the \textit{additional enthalpy} that is introduced for each structure as a result of cation disorder. This is the change in enthalpy as a result of geometric distortions and high energy bonds induced by disorder, rather than entropic effects. Thus, we have shown that in the \ce{ZnZrN2} polymorph structures with inequivalent cation sites (WS, LC), cation disordering significantly increases enthalpy, whereas in the \ce{ZnZrN2} structures with equivalent cation sites (RS, BN) cation disordering only negligibly increases enthalpy.

Four computational methods and resulting sets of energies have been considered so far: DFT to compute formation energies of ordered structures in \autoref{fig:prototypes}(b) ($\Delta H_{k}^\mathrm{ord}$), DFPT to estimate vibrational contributions in the SM ($G_k^\mathrm{vib}$), SQS to estimate random disordered structures in \autoref{fig:sqs}(b) ($\Delta H_{k}^\mathrm{dis}$), and the polymorph sampler ensemble to model configurational degrees of freedom in \autoref{fig:free_energy}(b) ($G_k^\mathrm{conf}$). Since an SQS structure approximates configurational disorder, it is also representative of a disordered structure that might be observed at high $T_\mathrm{eff}$. Thus, the $\Delta H_{k}^\mathrm{dis}$ represent similar disordered structures as the $G_k^\mathrm{conf}$ at sufficiently high $T_\mathrm{eff}$ (e.g. at $T_\mathrm{eff}$ = 2000 K, chosen as a representative high $T_\mathrm{eff}$ as depicted in \autoref{fig:free_energy}), with the former accessing enthalpy and the latter accessing enthalpy and entropy. \autoref{fig:comparison} shows that the relative polymorph ordering changes across the four methods: WS is lowest in $\Delta H_{k}^\mathrm{ord}$ and $G_k^\mathrm{vib}$ (at "high" temperatures of $T$ = 630 K), disordered BN is lowest in $\Delta H_{k}^\mathrm{dis}$, and RS is lowest in $G_k^\mathrm{conf}$ (at "high" effective temperatures of $T_\mathrm{eff}$ = 2000 K).

Since $G_k^\mathrm{conf}$ includes both enthalpic \textit{and} entropic degrees of freedom, and $\Delta H_{k}^\mathrm{dis}$ includes only enthalpic degrees of freedom, comparison of \autoref{fig:comparison}(c) and (d) suggests that at high disorder the RS phase is \textit{entropically stabilized} compared to BN. Meanwhile, at high disorder the WS phase is \textit{enthalpically destabilized} compared to the RS and BN phase. We reiterate that vibrational effects as shown in (b) do not induce significant energetic reordering. Distorted structures (d-RS and d-BN) are excluded from \autoref{fig:comparison} for clarity, though none of them are the lowest energy structure in any of these calculations. We acknowledge that the SQS disordered structures that determine $\Delta H_{k}^\mathrm{dis}$ are estimates (for example, a cluster expansion could be fit to rigorously account for short-range order and obtain a more accurate estimate), but ultimately these estimates support the hypothesis that disorder tolerance stabilizes the RS and BN phases over other considered polymorphs at high effective temperatures. Additionally, we have distinguished which stabilization effects are due to changes in enthalpy and which are due to changes in configurational entropy.

\subsection{Tolerance to off-stoichiometry informs phase transitions}

It has been suggested why RS and BN are stabilized with sputtering, but it is not yet understood why RS is synthesized at Zr-rich compositions and BN at Zn-rich compositions. Another plausible explanation for the absence of WS \ce{ZnZrN2} is that it is a "line compound," a phenomena observed in other ternary nitride systems such as \ce{ZnSnN2}:\ce{ZnO}.\cite{pan2020perfect} A line compound is stable only in a very narrow region of configurational space such that it may be missed using combinatorial growth.

\begin{figure}
    \centering
    \includegraphics[width=75 mm]{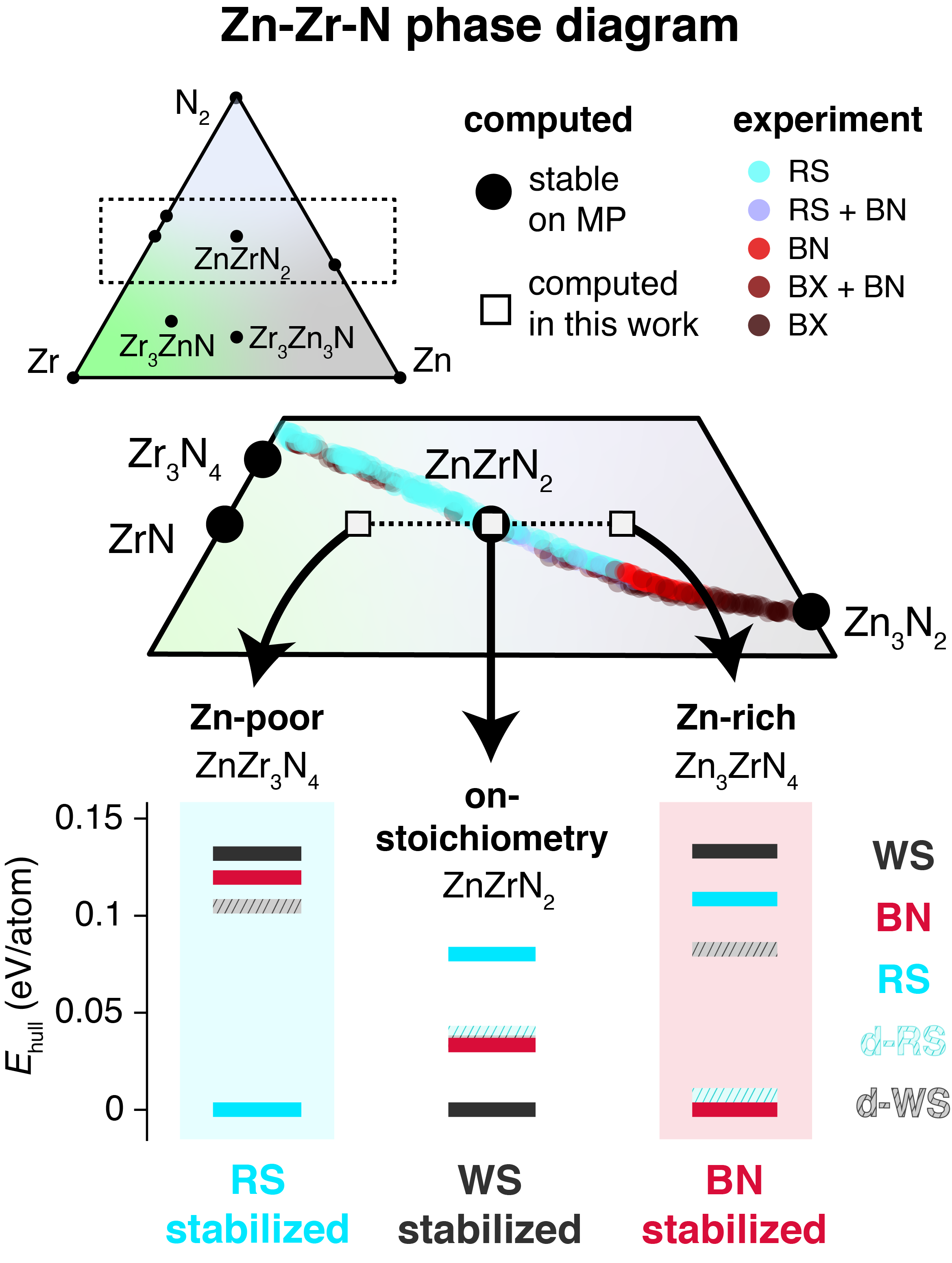}
    \caption{Ternary phase diagram of Zn-Zr-N and a close-up of the \ce{Zn_{$x$}Zr_{1-$x$}N} region, with computed compounds from the Materials Project designated. Colored circles depict the approximate composition and associated phases of experimental data from this work (see \autoref{fig:phase_map}). Computed compositions from this study are represented with unfilled squares in the phase diagram, and for each composition the lowest formation energy structure for a given class is plotted with colored bars.}
    \label{fig:alloy}
\end{figure}

\textbf{\autoref{fig:alloy}} plots ternary Zn-Zr-N phase space, which constitutes configurational space in this system, and shows approximately where experimental samples lie with respect to computed phases. To explore this hypothesis, we perform cation substitution in each of the ordered \ce{ZnZrN2} polymorph structures from \autoref{fig:prototypes} to create a set of prototypes across the ZrN--ZnN tieline --- i.e. \ce{Zn_{$x$}Zr_{1-$x$}N_{$y$}} where $y$ = 1 and $x$ = 0.25, 0.50, 0.75 ($x$ = 0 in SM) --- and then relax the structures (note that these are small ordered unit cells, not SQS cells). This constitutes a very rough alloy approximation, since sputtered films are N-rich for $x$ < 0.5 and N-poor for $x$ > 0.5; the experimental alloy is closer to the \ce{Zr3N4}--\ce{Zn3N2} tieline, but this heterovalent alloy is far trickier to model due to defect compensation and is not performed here. 

Relative polymorph stability for relevant structure classes with $E_\mathrm{hull}$ < 0.15 eV/atom is depicted for \ce{ZnZr3N4}, \ce{ZnZrN2}, and \ce{Zn3ZrN4} in \autoref{fig:alloy} (see SM for hull stability plots and all classes). WS is highly destabilized in Zn-poor and Zn-rich cases, suggestive of a line compound. RS is the lowest energy polymorph for a wide window of Zn-poor compositions and BN is the lowest energy polymorph in Zn-rich compositions (see SM). Since the LC, ZB, and WZ phases do not emerge in the polymorph sampler and given that WS is highly metastable at high $T_\mathrm{eff}$ (see \autoref{fig:free_energy}), at high $T_\mathrm{eff}$ we would expect a phase change from RS to BN somewhere within approximately 0.45 < $x$ < 0.55 (see SM), which corroborates experimental findings. These calculations use the nominal valence of the cations, namely, Zn\textsuperscript{2+} and Zr\textsuperscript{4+}; we do not perform defect calculations nor vary cation oxidation states. Rigorous examination of alloy phase space would require an in-depth calculation of a temperature-dependent phase diagram, which is beyond our scope. However, this simple approximation is sufficient to explain our experimental observation of a phase change from RS to BN as $x$ increases in \ce{Zn_{$x$}Zr_{1-$x$}N_{$y$}} at high $T_\mathrm{eff}$. This is supported by the previous discussion on disorder tolerance: in order to achieve off-stoichiometry, cations have to be placed on energetically unfavorable sites.

\subsection{Implications for materials discovery}

Although neglected in this analysis, it is important to acknowledge the role of dynamic, kinetic, and additional entropic effects in this ternary phase space. Electronic contributions to entropy have shown to be negligible in solid alloys.\cite{manzoor2018entropy} As has been demonstrated in other II-IV-\ce{N2} systems, spurious oxygen incorporation from the growth chamber can influence phase stability and result in impurity phases,\cite{greenaway2020combinatorial} though our O/(Zn+Zr) values below 1\% from \autoref{fig:phase_map}(a) suggest that a phase-segregated oxide is not observed here. Additionally, the elemental Zr used in our sputter target contains $\sim$10 at. \% Hf, which could influence relative phase stability. Finally, surface morphology, templating during growth, and kinetic effects could be assisting in restricting the formation of WS, and in enabling the preferential formation of RS and BN. Even though these films are grown on amorphous fused silica, we also acknowledge the possibility of preferential nucleation. These factors are all important to take into account in materials discovery studies.


\begin{figure*}
    \centering
    \includegraphics[width=0.8\textwidth]{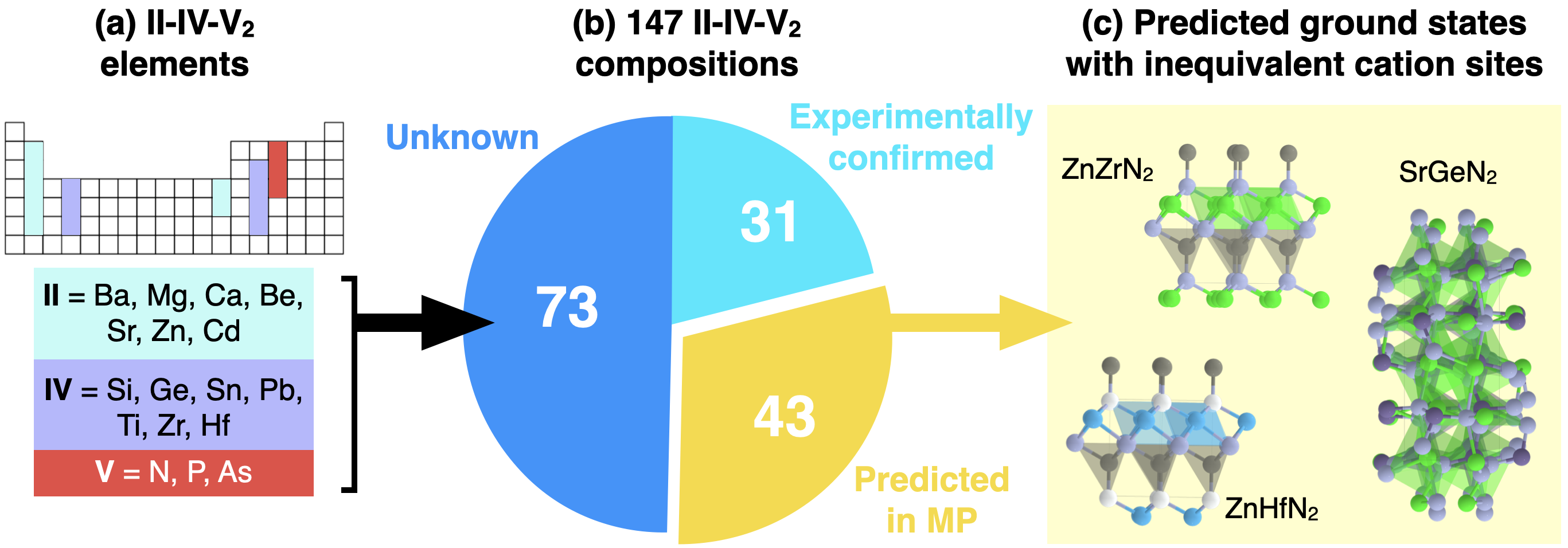}
    \caption{Assessment of the role of cation disorder tolerance in emerging ternary pnictides. (a) Elements within the II-IV-V\ce{_2} composition space. (b) The distribution of composition spaces within the Materials Project (MP) database. (c) Three example systems from the "predicted in MP" category where cations occupy symmetrically inequivalent lattice sites.}
    \label{fig:thought-exp}
\end{figure*}

There are many new predicted ternary nitrides and pnictides to explore beyond \ce{ZnZrN2}.\cite{sun2019map} However understanding of which polymorphs are actually synthesizable remains elusive. An assessment of the role of cation disorder tolerance in emerging ternary pnictide systems is presented in \textbf{\autoref{fig:thought-exp}}, with the set of II-IV-V\ce{_2} pnictide semiconductors considered in (a) where II = (Ba, Mg, Ca, Be, Sr, Zn, Cd), IV = (Si, Ge, Sn, Pb, Ti, Zr, Hf), and V = (N, P, As). Theoretically, this set includes 147 unique compositions; as shown in (b), to date only 31 of these compositions have been confirmed experimentally and only 43 other predicted compositions are in the Materials Project (MP) database, leaving 73 II-IV-V\ce{_2} compositions not yet on the database at this time. Within the set of predicted but not-yet-synthesized compounds (yellow), (c) shows that at least two other systems other than \ce{ZnZrN2} have computed ground states where cations occupy symmetrically inequivalent lattice sites: \ce{SrGeN2} ($Pbca$) and \ce{ZnHfN2} ($P3m1$). This is also feasible for the 73 ternary pnictide compositions still missing from the MP database, leading to the experimental realizability of new metastable compounds with promising properties to be explored.

\section{Conclusion}

In summary, we have grown the first \ce{Zn_{$x$}Zr_{1-{$x$}}N_{$y$}} samples (a set of approximately 400 thin films) using combinatorial sputter synthesis, demonstrating the crystallization of metastable rocksalt-derived (RS) and boron-nitride-derived (BN) phases with cation-disordered structures, rather than the predicted cation-ordered "wurtsalt" (WS) ground state phase. These findings have been explained using first-principles computational methods. By statistically sampling configurational degrees of freedom of polymorphs generated by random structure sampling, we demonstrate energetic destabilization of the predicted DFT WS ground state at high temperatures and stabilization of RS and BN phases that support our experimental results. It is shown that this stabilization can be attributed to the increased disorder tolerance in the RS and BN phases due to only minor gains in configurational enthalpy, suggesting that the RS phase is entropically stabilized to a higher degree than BN. Ordered alloy calculations of varying cation composition suggest that RS and BN have a higher tolerance to cation off-stoichiometries compared to WS, predicting a phase transformation from RS to BN as $x$ increases that corroborates our experimental findings. These results demonstrate the utility of sputtering in accessing high effective temperatures and synthesizing polymorphs predicted to be metastable within the DFT approximation at 0 K. 

However, we acknowledge that growth methods and deposition conditions matter significantly in phase stabilization, and sputtering of thin films is just one synthesis approach. Although WS is energetically destabilized here by sputter synthesis, its realizability is not definitively ruled out. Future work on targeted synthesis of WS phases (e.g. low effective temperature, epitaxial, on-stoichiometry synthesis) is needed to assess whether WS is indeed synthesizable. If synthesizable, WS \ce{ZnZrN2} holds promise as a piezoelectric material and for optoelectronic applications.\cite{tholander2016strong, ling2020origin} Furthermore, in-depth structural analysis and optoelectronic properties of the RS and BN polymorphs in this system remain to be investigated. \autoref{tab:polymorph_energies} indicates promising properties for device applications such as contact materials, solar cell absorbers, photocatalysts, piezoelectric and ferroelectric materials.\cite{tholander2016strong, ling2020origin} In particular, the synthesized BN-derived polymorph has $>$2 eV band gap and low (<1.5) well-matched electron and hole effective masses, making it interesting for electronic devices that can operate at elevated temperatures. Additionally, this non-polar BN polymorph is the transition state between two variants of the polar WZ structure, suggesting a pathway to tuning its predicted and measured ferroelectric response.\cite{dreyer2016correct, fichtner2019alscn}

The results of this work suggest that other thermodynamically "metastable" materials according to 0 K DFT may be possible to synthesize. Presently DFT is one of the most popular methods to generate high-throughput thermochemistry data with reasonably accuracy, despite the fact that zero temperature formation energies provide only a rough estimate of actual material stability. In extended inorganic solids, a general rule-of-thumb is that entropy contributes on the order of $\sim$0.05--0.1 eV/atom to the free energy. Accordingly, many high-throughput computational screening studies discard materials that have an $E_\mathrm{hull}$ above a cutoff of $\sim$0.05--0.1 eV/atom. However, this study demonstrates synthesis of a RS polymorph phase with $E_\mathrm{hull}$ in the range of $\sim$0.08--0.15 eV/atom using a common PVD technique, suggesting stabilization due to disorder tolerance. Since this phase would have been ruled out as not-synthesizable by a typical high-throughput computational screening, it may be important to revisit what other metastable but synthesizable phases have been overlooked in such studies.

A contemporary challenge in materials science research is to bridge the gap between computationally predicted materials and experimental materials that can actually be grown in the laboratory with desired properties. The \ce{ZnZrN2} results presented in this study are interesting beyond this specific material system because there may be many accessible energetic states that neither scientists nor nature have realized yet. In the Zn-Zr-N material system, it appears that tolerance to disorder and off-stoichiometry contribute to the realization of high formation energy phases, and this study has introduced a new descriptor to assess disorder tolerance, $E_\mathrm{rdp}$. However, in other material systems there may be different physical mechanisms enabling synthesis of metastable polymorphs. In general, the computational materials discovery community needs to continue to redefine the metrics by which phase stability and synthesizability are assessed in order to yield experimentally realizable predictions that enable new functional materials.

\section*{Acknowledgments}

This work was authored in part at the National Renewable Energy Laboratory, operated by Alliance for Sustainable Energy, LLC, for the U.S. Department of Energy (DOE) under Contract No. DE-AC36-08GO28308. Funding was provided by the Office of Science (SC), Office of Basic Energy Sciences (BES), Materials Chemistry program, as a part of the Early Career Award “Kinetic Synthesis of Metastable Nitrides”. R.W.R. acknowledges financial support from the U.C. Berkeley Chancellor's Fellowship and the National Science Foundation (NSF) Graduate Research Fellowship under Grant No. DGE1106400 and DGE175814. V.S. acknowledges financial support from NSF Career Award No. DMR-1945010 for polymorph sampler ensemble calculations. Use of the Stanford Synchrotron Radiation Lightsource, SLAC National Accelerator Laboratory, is supported by DOE's Office of Science (SC), Basic Energy Sciences (BES) under Contract No. DE-AC02-76SF00515. The computational work was supported by the U.S. Department of Energy, Office of Science, Office of Basic Energy Sciences, Materials Sciences and Engineering Division under Contract No. DE-AC02-05-CH11231 (Materials Project program KC23MP). The authors thank Dr. Sage Bauers, Dr. Kevin Talley, Valerie Jacobson, and Rachel Sherbondy for experimental assistance, Dr. Shyam Dwaraknath, Eric Sivonxay, and Matthew McDermott for computational assistance, Dr. John Perkins and Dr. Apurva Mehta with characterization assistance, and Dr. Adele Tamboli and Dr. Wenhao Sun for insightful discussions. This work used high-performance computing resources located at NREL and sponsored by the Office of Energy Efficiency and Renewable Energy. The views expressed in the article do not necessarily represent the views of the DOE or the U.S. Government.

\section*{Supplementary materials}

The following supplemental material (SM) is included:

\begin{itemize}
  \item S1 Synthesis details
  \item S2 Characterization details
  \item S3 Polymorph and DFT formation energy ordering
  \item S4 Computational methods details
\end{itemize}


\section*{Author contributions}

Conceptualization, R.W.R., A.Z., K.A.P.; Methodology, R.W.R., V.S., S.L., K.N.H.; Computational Investigation, R.W.R., V.S., S.L., M.K.H.; Experimental Investigation, R.W.R., K.N.H., A.Z.; Writing - Original Draft, R.W.R., V.S.; Writing – Review \& Editing, R.W.R., A.Z., S.L., M.K.H., K.A.P.; Funding Acquisition, R.W.R., A.Z., V.S., K.A.P.; Supervision, A.Z., V.S., K.A.P.

\section*{Data availability statement}

All ordered crystal structures will be uploaded to the Materials Project database and available free of charge. Experimental data is available on the High Throughput Experimental Materials (HTEM) Database.\cite{zakutayev2018open} All data needed to evaluate the conclusions in the paper are present in the paper and/or the Supplementary Materials.

\bibliographystyle{ieeetr}
\bibliography{refs.bib}

\end{document}



\title{Supplemental Material for "The role of disorder in the synthesis of metastable zinc zirconium nitrides"}




    
\author{Rachel Woods-Robinson\textsuperscript{*,1,2,3}, Vladan Stevanović\textsuperscript{4,3}, Stephan Lany\textsuperscript{3}, Karen N. Heinselman\textsuperscript{3}, Matthew K. Horton\textsuperscript{2}, Kristin A. Persson\textsuperscript{5,2}, Andriy Zakutayev\textsuperscript{*,3}}

\affiliation{\textsuperscript{1}Applied Science and Technology Graduate Group, University of California at Berkeley, Berkeley, CA, 94720 USA, \textsuperscript{2}Materials Sciences Division, Lawrence Berkeley National Laboratory, Berkeley, CA, 94720 USA, \textsuperscript{3}Materials Science Center, National Renewable Energy Laboratory, Golden, Colorado, 80401 USA \textsuperscript{4}Department of Physics, Colorado School of Mines, Golden, Colorado, 80401 USA,
\textsuperscript{5}Department of Materials Science and Engineering, University of California at Berkeley, Berkeley, CA, 94720 USA}

\date{\today}

\maketitle

\section{Synthesis details}

\begin{figure*}[!htb]
    \centering
    \includegraphics[width=160mm]{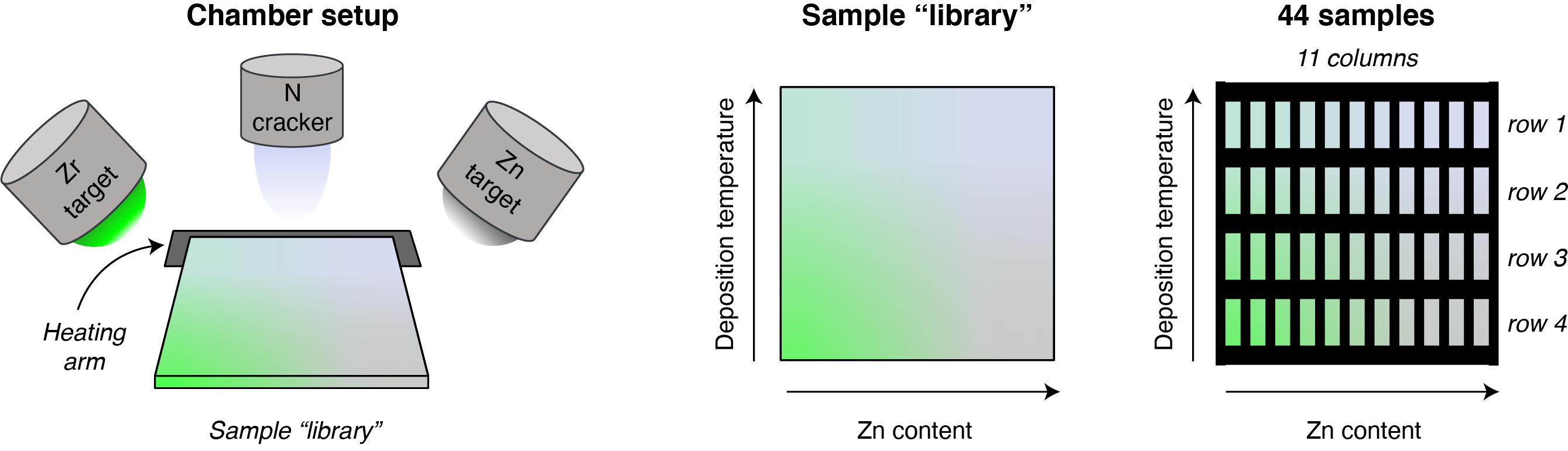}
    \caption{Combinatorial experimental setup.}
    \label{fig-si:combi_setup}
\end{figure*}

\subsection{Combinatorial synthesis details}

The combinatorial sputter chamber setup is depicted in \textbf{Figure \ref{fig-si:combi_setup}}. Growth conditions for samples discussed in this manuscript are reported in \textbf{Table \ref{tab-si:dep_conditions}}. To estimate deposition temperature $T_\mathrm{dep}$ across the combinatorial library as reported in Figure 2, we adopt the following procedure. A gradient in growth temperature is induced perpendicular to the Zn content gradient, as shown in Figure \ref{fig-si:combi_setup}, by contacting the substrate on the top end with a heated metal pad, as described by Subramaniyan et al.\cite{subramaniyan2014non} The induced temperature gradient has been previously calibrated by Fioretti et al.,\cite{fioretti2015combinatorial} and by other researchers in our group. It is estimated from the substrate temperature using an exponential decay function such that the $T_\mathrm{dep}$ of the first row is approximately equal to the set-point temperature $T_\mathrm{set}$ (as reported in Table \ref{tab-si:dep_conditions}). The $T_\mathrm{set}$ is varied from ambient to 500\degree C, though only samples with $T_\mathrm{set}$ up to 310\degree C are reported here, as explained subsequently. We note the uncertainty in quantitative temperature as a result of this approximation, and as a result of non-uniform substrate heating, but qualitative trends hold.

\begin{table*}[]
\caption{Deposition conditions and end-point compositions for the 11 libraries reported here.}
\label{tab-si:dep_conditions}
\resizebox{\textwidth}{!}{%
\begin{tabular}{@{}cccccccccccc@{}}
\toprule
\textbf{\begin{tabular}[c]{@{}c@{}}Sample\\ number\end{tabular}} &
  \textbf{\begin{tabular}[c]{@{}c@{}}Time\\ (min)\end{tabular}} &
  \textbf{\begin{tabular}[c]{@{}c@{}}Zn:Zr power\\ ratio\end{tabular}} &
  \textbf{\begin{tabular}[c]{@{}c@{}}Zn target\\ power (W)\end{tabular}} &
  \textbf{\begin{tabular}[c]{@{}c@{}}Zr target\\ power (W)\end{tabular}} &
  \textbf{\begin{tabular}[c]{@{}c@{}}$\bm{T_\mathrm{set}}$\\ (C) \end{tabular}} &
  \textbf{\begin{tabular}[c]{@{}c@{}}Calib. $\bm{T_\mathrm{dep}}$\\ (C) (row 1)\end{tabular}} &
  \textbf{\begin{tabular}[c]{@{}c@{}}Min. $\bm{x}$\\ (row 1)\end{tabular}} &
  \textbf{\begin{tabular}[c]{@{}c@{}}Max. $\bm{x}$\\ (row 1)\end{tabular}} &
  \textbf{\begin{tabular}[c]{@{}c@{}}Calib. $\bm{T_\mathrm{dep}}$\\ (C) (row 4)\end{tabular}} &
  \textbf{\begin{tabular}[c]{@{}c@{}}Min. $\bm{x}$\\ (row 4)\end{tabular}} &
  \textbf{\begin{tabular}[c]{@{}c@{}}Max. $\bm{x}$\\ (row 4)\end{tabular}} \\ \midrule
820 & 180 & 0.67 & 40  & 60 & 220 & 275 & 0.038 & 0.072 & 180 & 0.285 & 0.358 \\
821 & 180 & 1.33 & 80  & 60 & 220 & 275 & 0.122 & 0.275 & 180 & 0.706 & 0.940 \\
822 & 150 & 2.00 & 80  & 40 & 310 & 350 & 0.015 & 0.033 & 250 & 0.312 & 0.405 \\
823 & 120 & 0.38 & 30  & 80 & -- & 60  & 0.503 & 0.728 & 60  & 0.613 & 0.819 \\
824 & 120 & 0.38 & 30  & 80 & -- & 60  & 0.100 & 0.342 & 60  & 0.119 & 0.461 \\
825 & 210 & 5.00 & 100 & 20 & 310 & 350 & 0.035 & 0.069 & 250 & 0.921 & 0.972 \\
826 & 120 & 0.38 & 30  & 80 & -- & 60  & 0.336 & 0.710 & 60  & 0.340 & 0.756 \\
827 & 113 & 0.00 & 0   & 80 & -- & 60  & 0.000 & 0.000 & 60  & 0.000 & 0.000 \\
828 & 66  & inf  & 80  & 0  & -- & 60  & 1.000 & 1.000 & 60  & 1.000 & 1.000 \\
829 & 180 & 1.00 & 60  & 60 & 130 & 200 & 0.371 & 0.696 & 100 & 0.578 & 0.901 \\
831 & 180 & 0.75 & 60  & 80 & 130 & 200 & 0.358 & 0.468 & 100 & 0.653 & 0.858 \\ \bottomrule
\end{tabular}}
\end{table*}

\subsection{A comment on high temperature synthesis}

At low to moderate temperatures, the power of the Zr gun has to be higher than that of the Zn gun to achieve stoichiometries of approximately $x$ = 0.5. As temperature increases, the Zn:Zr power ratio has to increase to maintain the same stoichiometry. This explains the range of growth conditions in the experimental figure in the manuscript. We attempted to grow ternary films higher than 400\degree C, but all films are Zr-rich and phase segregated. This is likely due to the low vapor pressure of Zn compared to Zr, which results in a lower tendency to stick as temperature increases. The gun power cannot be increased higher than 100 W for safety reasons, so to attempt to counter this we installed a second Zn target into the chamber. Setting two Zn guns to 100 W (maximum power) gave an effective Zn power of 200 W, and lowering the Zr gun to 20 W gave a Zn:Zr power ratio of 10:1; however, this was still not high enough for Zn to stick. Therefore, we only report results at temperatures below calibrated $T_\mathrm{dep}$ values of 350\degree C.

\subsection{Annealing}

We conducted a series of annealing experiments of ambient-temperature deposited \ce{Zn_{$x$}Zr_{1-$x$}N} thin film samples. Anneals were performed on combinatorial library rows in an \ce{N2} environment, with a one minute ramp-up and ramp-down time, at temperatures ranging from 100--900\degree C. Lower temperatures ($\sim$100\degree C) do not change the composition or diffraction peaks. However, upon annealing at temperatures higher than $\sim$300\degree C, nearly all Zn in the films is lost, yielding Zr-rich rocksalt-derived crystals. We tried two depositions with AlN capping layers, followed by an anneal, but films appeared to phase segregate into binary \ce{ZrN} and \ce{Zn3N2}, and results were inconclusive. However, as our exploration was not exhaustive, we suggest that followup experiments using a capping layer and annealing may be useful to pursue in attempts to synthesize the predicted "wurtsalt" phase.

\subsection{A comment on deposition temperatures vs. effective temperatures}

We note that it is very common that for oxides or nitrides up to $T_\mathrm{dep}$ of at least 400--500\degree C, the “effective temperature” $T_\mathrm{eff}$ increases with decreasing  substrate temperature. This is the temperature that is referenced in calculations. This is because at low substrate temperature adatoms do not have enough energy to move around on the substrate and get essentially "frozen" into a non-equilibrium disordered position. This trend can be reversed at higher substrate temperatures (above at least 600--800 \degree C), where adatoms do have enough energy, but this is beyond the $T_\mathrm{dep}$ range explored in this study.

\section{Composition and structural characterization}

\subsection{A pedantic comment on ternary nitride alloy notation}

Within the field of emerging ternary nitrides, there is current no "best practice" on whether to notate the off-stoichiometric spaces of new materials in \textit{alloy} notation, which emphasizes the variation of a ternary system between two binaries, or \textit{compound} notation, which emphasizes the off-stiochiometry from a particular ternary compound phase. For this system, alloy notation would look like e.g. "\ce{Zn_$x$Zr_{1-$x$}N_$y$}" and compound notation like e.g. "\ce{Zn_{1+$x$}Zr_{1-$x$}N_{2-$y$}}".

On one hand, compound notation \ce{Zn_{1+$x$}Zr_{1-$x$}N_{2-$y$}} would emphasize \ce{ZnZrN2} as the on-stoichiometric compound, and “$x$” as any variation from this stoichiometry. This could be advantageous in designating compounds rather than binary alloys. This notation is used across the literature, and in nitrides specifically such as \ce{ZnSnN2} and \ce{ZnGeN2} that have been demonstrated to be line compounds. On the other hand, alloy notation \ce{Zn_$x$Zr_{1-$x$}N_$y$} is more general. The alloy notation is common for isovalent alloys, where one can change the cation ratio without affecting the anion composition. Thus, much of the ternary nitride literature uses “\ce{A_$x$B_{1-$x$}C}” (e.g. \ce{Cr_$x$Mo_{1-$x$}N_$y$} or \ce{B_$x$Al_{1-$x$}N_$y$}), though most of these have to do with elemental substitution in the binary WZ or RS structure (e.g. Mo in RS-CrN or B in WZ-AlN) rather than a unique ternary compound. More recently, this notation has been adapted to refer to systems where ternary nitrides occur, e.g. \ce{Mg_{$x$}$TM$_{1-$x$}N_{$y$}}, and can be updated for aliovalent compounds by changing $y$.

Ideally in this Zn-Zr-N study it would be useful to use compound notation to emphasize the \ce{ZnZrN2} stoichiometry, but compound notation presents a few challenges. First, this would necessitate using negative values of "$x$" for anything Zr-rich, which is not common practice in the literature to our knowledge. Second, this study shows, as the Zn/(Zn+Zr) ratio increases, a phase shift occurs from RS to BN to BX \ce{Zn3N2}. Thus there's a RS alloy up to a point (Zn/(Zn+Zr) $\approx$ 0.66), so \ce{Zn_$x$Zr_{1-$x$}N_$y$} alloy notation is more accurate at least for Zr-rich compositions. Most importantly, \ce{Zn_{1+$x$}Zr_{1-$x$}N_{2-$y$}} \textit{a priori} assumes there is a RS or BN compound \textit{exactly} at the \ce{ZnZrN2} stoichiometry, but we don’t actually know this or demonstrate this---the actual BN “compound” may lie at a more Zn-rich composition such as \ce{Zn3ZrN4} (or, e.g. could be a topotactic alloy of a \ce{ZnN_$y$} polymorph phase with Zr insertion). Thus, for this study we select alloy notation \ce{Zn_$x$Zr_{1-$x$}N_$y$} because it is more general and because the investigation spans multiple phases.

\subsection{Composition analysis}

\begin{figure}
    \centering
    \includegraphics[width=80mm]{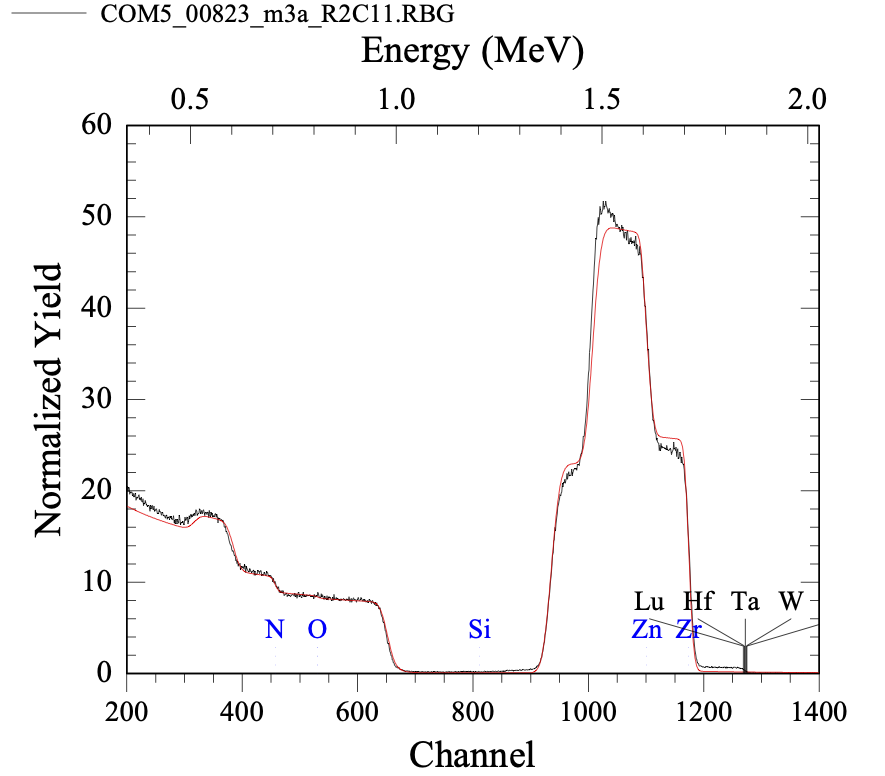}
    \caption{RBS measurement example.}
    \label{fig-si:rbs}
\end{figure}

An example of the RBS data from sample 823 is reported in \textbf{Figure \ref{fig-si:rbs}}. We note that RBS is performed only on samples deposited at ambient temperature, and that it is possible that the anion concentration changes at higher growth temperatures for a given cation ratio. Using RBS, we can estimate how the concentration of nitrogen changes across the cation alloy space in our samples. Additionally, oxygen is ubiquitous and nearly unavoidable in nitride semiconductors -- specifically oxygen has been shown to contaminate ternary nitride materials grown in the same chamber (e.g. LaWON). RBS can assess the extent of oxygen contamination in these films, within an accuracy of approximately 1\%.

\subsection{Structural analysis}

\subsubsection{X-ray diffraction details}

\begin{figure*}
    \centering
    \includegraphics[width=\textwidth]{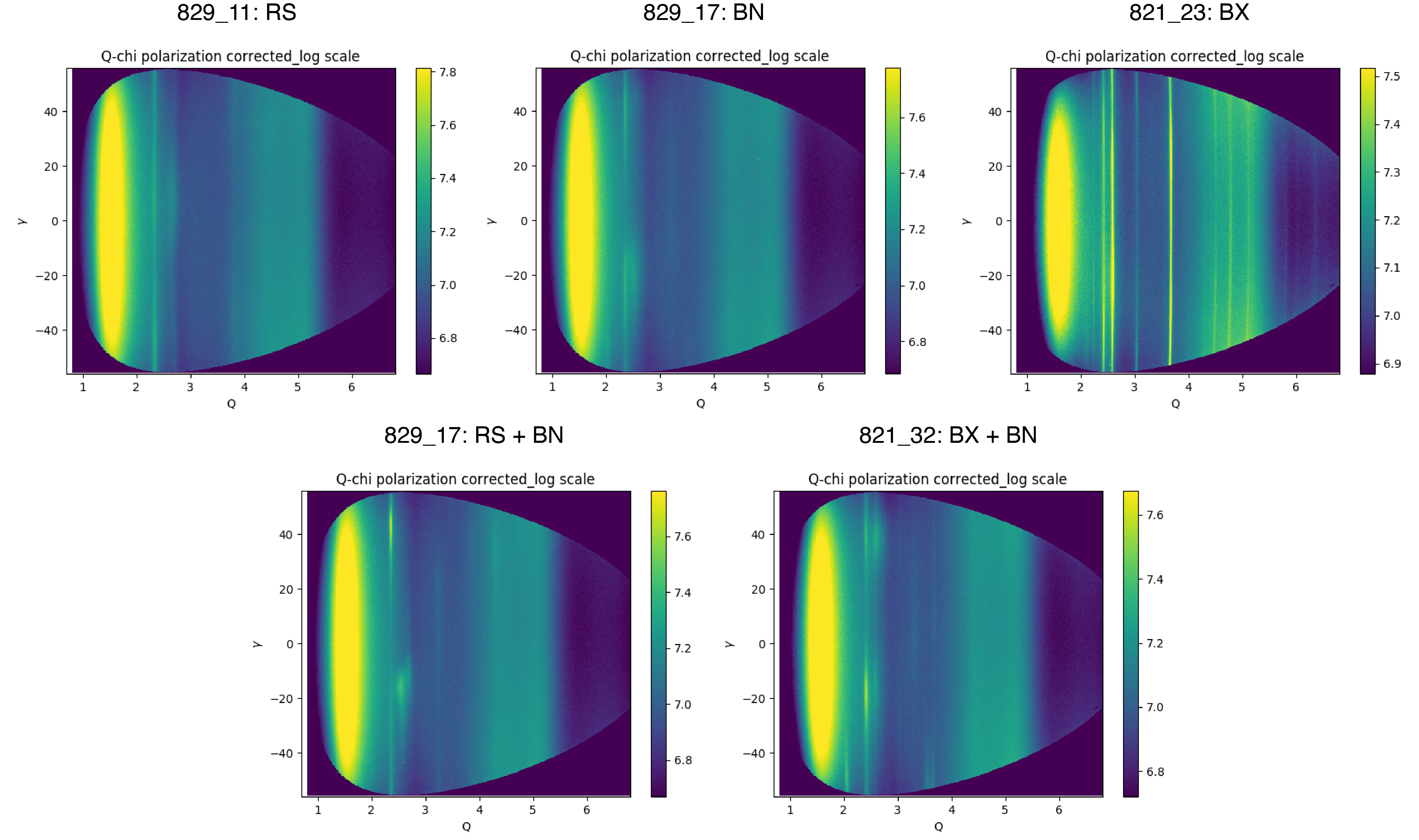}
    \caption{Synchrotron WAXS data plotted as $Q$ vs. $\gamma$ for five representative samples.}
    \label{fig-si:q_chi}
\end{figure*}

$Q$ vs. $\gamma$ plots are given in \textbf{Figure \ref{fig-si:q_chi}} for reference, depicting the calibrated data from the synchrotron and the texturing within films. Note that our data set consists of over 400 of such plots. We have not analysed texturing here, but provide the data so that other researchers may investigate if interested. To create the $Q$ vs. intensity data reported in the manuscript, $Q$ vs. $\gamma$ data was integrated between 5\degree<$\gamma$<175\degree.

\subsubsection{Peak identification and phase shifts}

\begin{figure*}
    \centering
    \includegraphics[width=\textwidth]{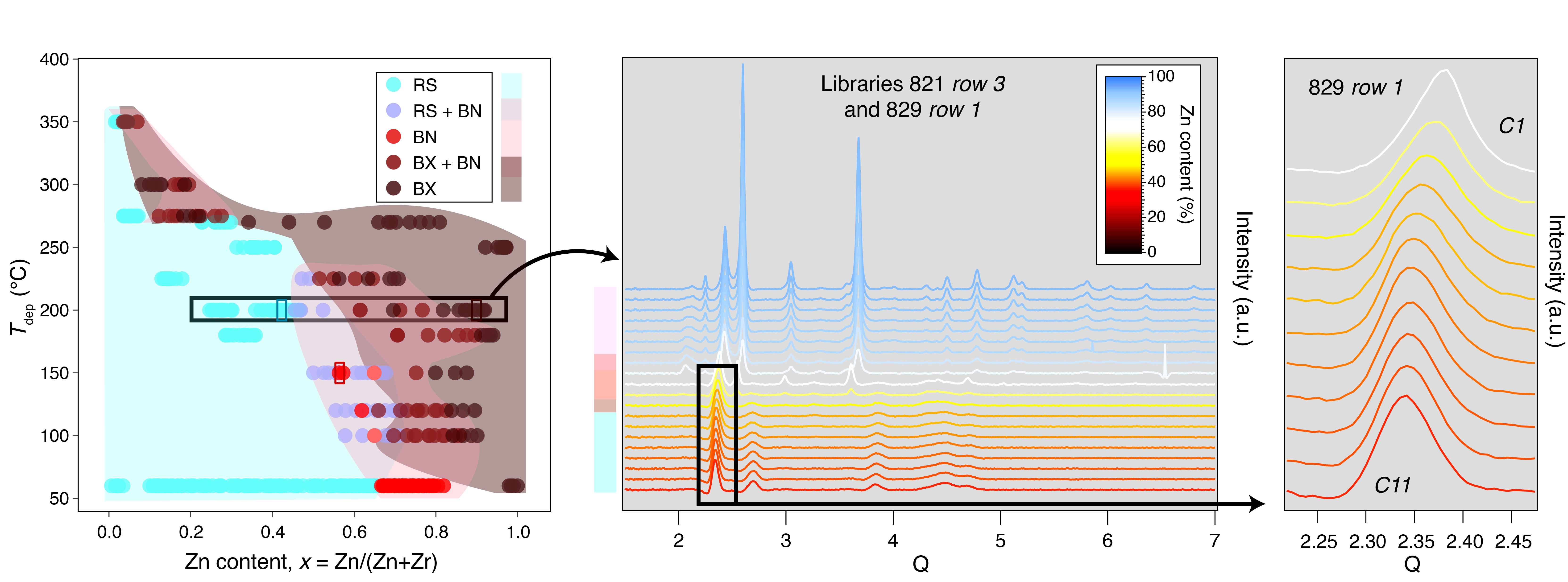}
    \caption{Diffraction patterns across composition space at approximately 200\degree C.}
    \label{fig-si:ssrl}
\end{figure*}

As reported in the manuscript, \ce{Zn_{$x$}Zr_{1-$x$}N} appear to crystallize in the RS, BN, and BX structure, as well as mixed phases. Although only three diffraction patterns are reported in the manuscript for brevity, and we are primarily interested in the $x$ = 0.5 region, we have collected over 400 diffraction patterns in this analysis. The manuscript reports an approximate map of phase space based on diffraction patterns. One of the key advantages of combinatorial data analysis is the ability to look at trends across composition and temperature phase space. We can also observe peak shifts as a function of cation composition from the combinatorial data, as depicted in \textbf{Figure \ref{fig-si:ssrl}}. The center panel of this figure shows, for a particular deposition temperature of approximately 200\degree C, the WAXS patterns corresponding to phase changes between RS, BN, and BX as the Zn content increases from 0\% to 100\%.

We can zoom in on the RS region (right-most panel) to the (111) reflections to observe how much the peak shifts as a function of composition. As $x$ increases, this peak shifts to higher $Q$ values. The reason for this shift could be due to Zn substituting on Zr sites in RS ZrN, but could also be due to anion change (see RBS results) and thus further analysis here is beyond our scope. Note that at some point this peak represents the BN (121) reflection, which could also contribute to the peak shift, and the nature of the phase transformation is not studied here. In contrast, there is no noticeable shift in the peaks of the BX patterns at this temperature, which is what would be expected if Zr was substituting for Zn on in \ce{Zn3N2}. Thus, it is plausible that this composition consists of BX \ce{Zn3N2} and a minor secondary phase of Zr-N, though such analysis is beyond our scope.

In the manuscript, three representative patterns are selected for plotting:

\begin{itemize}
    \item RS: $x$ = 0.45, $T_\mathrm{dep}$ = 200\degree C (library 831 sample \#6 and library 829 sample \#4)
    \item BN: $x$ = 0.56, $T_\mathrm{dep}$ = 150\degree C (library 829 sample \#18 and \#19)
    \item BX: $x$ = 0.92, $T_\mathrm{dep}$ = 200\degree C (library 821 sample \#23, library 829 sample \#23 and \#24)
\end{itemize}

\section{Polymorph and DFT ordering}

\subsection{Comparing SCAN, PBE, and PBE+U}

\begin{figure*}
    \centering
    \includegraphics[width=160mm]{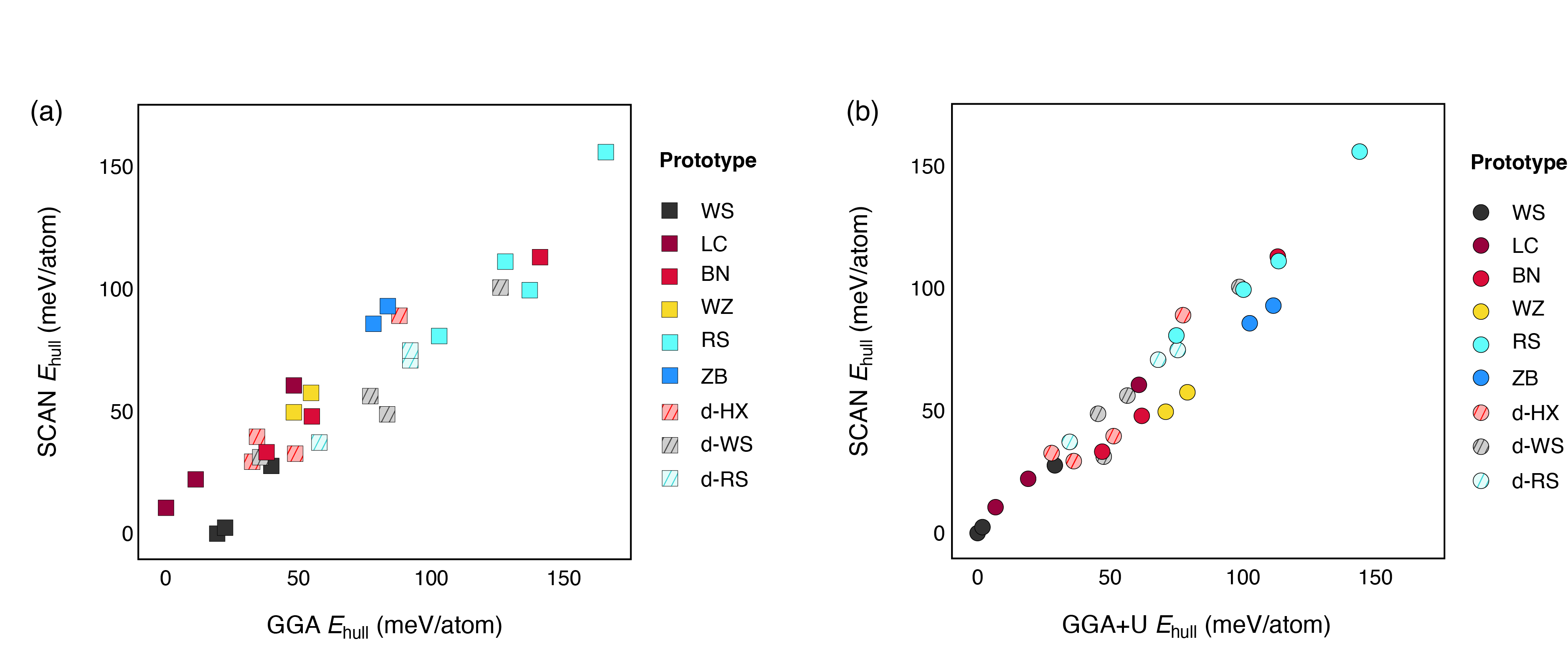}
    \caption{Comparison of formation energies of (a) SCAN vs. PBE and (b) SCAN vs. PBE+U, where U = 3 eV/atom on the Zr d orbitals.}
    \label{fig-si:SCAN_PBE}
\end{figure*}

For completeness, we calculate formation energies of the set of \ce{ZnZrN2} polymorphs with three different functional schemes: PBE, PBE+U (U = 3 eV/atom), and SCAN (see below for methodology). \textbf{Figure \ref{fig-si:SCAN_PBE}} compares the formation energies of these three functionals for a set of polymorphs, color coded by prototype class. It appears that including the Hubbard U value (3 eV/atom on Zr) is closest to the energies calculated with SCAN, however it is unclear whether 3 eV/atom is an appropriate choice of U.
Calculation results and energies for each structure (an extension of Table 1 in the manuscript) are also reported in \textbf{Table \ref{tab-si:SI_calcs}}.

\begin{table*}[!htb]
\caption[\ce{ZnZrN2} polymorph calculations]{\ce{ZnZrN2} polymorph calculations (note: $E_\mathrm{G}$, $E_\mathrm{G}^\mathrm{d}$, $m^*_\mathrm{e}$, and $m^*_\mathrm{h}$ calculated with SCAN).}
\label{tab-si:SI_calcs}
\resizebox{0.75\textwidth}{!}{
\begin{tabular}{@{}cccccccccccc@{}}
\toprule
\textbf{Prototype} &
  \textbf{Name} &
  \textbf{\begin{tabular}[c]{@{}c@{}}Space group\\ number\end{tabular}} &
  \textbf{\begin{tabular}[c]{@{}c@{}}Space group\\ symbol\end{tabular}} &
  \textbf{\begin{tabular}[c]{@{}c@{}}\# of\\ atoms\end{tabular}} &
  \textbf{\begin{tabular}[c]{@{}c@{}}PBE+U $\bm{E_\mathrm{hull}}$\\ (meV/at)\end{tabular}} &
  \textbf{\begin{tabular}[c]{@{}c@{}}SCAN $\bm{E_\mathrm{hull}}$\\  (meV/at)\end{tabular}} &
  \textbf{\begin{tabular}[c]{@{}c@{}}$\bm{E_\mathrm{G}}$ \\ (eV)\end{tabular}} &
  \textbf{\begin{tabular}[c]{@{}c@{}}$\bm{E_\mathrm{G}^\mathrm{d}}$ \\ (eV)\end{tabular}} &
  \textbf{$\bm{m^*_\mathrm{e}}$} &
  \textbf{$\bm{m^*_\mathrm{h}}$} &
  \textbf{\begin{tabular}[c]{@{}c@{}}In \\ Table 1\end{tabular}} \\ \midrule

WS   & ZZN-156  & 156 & $P3m1$     & 4  & 0    & 0     & 2.47 & 3.1  & 7.3  & 1.69 & * \\
WS   & ZZN-186  & 186 & $P6_3mc$   & 8  & 1.8  & 2.4   & 2.25 & 2.88 & 2.59 & 1.76 &   \\
LC   & ZZN-29   & 29  & $Pca2_1$   & 16 & 6.8  & 10.6  & 1.63 & 1.63 & 1.33 & 1.87 & * \\
LC   & ZZN-45   & 45  & $Iba2$     & 32 & 19.1 & 22.2  & 1.61 & 1.61 & 1.72 & 1.55 &   \\
WS   & ZZN-164  & 164 & $P\bar{3}m1$    & 8  & 29.1 & 27.7  & 0.77 & 1.37 & 1.97 & 2.27 &   \\
d-HX   & ZZN-14   & 14  & $P2_1/c$   & 16 & 36.2 & 29.4  & 2.62 & 2.71 & 3.88 & 2.16 & * \\
d-WS & ZZN-8a   & 8   & $Cm$       & 32 & 47.5 & 31.2  & 2.18 & 2.18 & 1.56 & 1.25 & * \\
BN   & ZZN-8    & 8   & $Cm$       & 16 & 27.9 & 32.7  & 2.01 & 2.01 & 1.36 & 1.49 & *  \\
BN   & ZZN-12   & 12  & $C2/m$     & 16 & 47   & 33.3  & 2.22 & 2.22 & 1.13 & 1.61 &   \\
d-RS & ZZN-7    & 7   & $Pc$       & 16 & 34.7 & 37.3  & 2.22 & 2.47 & 3.41 & 2.17 & *  \\
d-HX & ZZN-129  & 129 & $P4/nmm$   & 8  & 51.3 & 39.6  & 1.96 & 1.96 & 0.58 & 4.04 &  \\
BN   & ZZN-51   & 51  & $Pmma$     & 8  & 61.9 & 47.9  & 1.81 & 2.37 & 1.97 & 3.61 &   \\
d-WS & ZZN-160a & 1   & $P1$       & 24 & 38.6 & 38.2  & 1.9  & 2.08 & 1.23 & 6.27 &  \\
WZ   & ZZN-26   & 26  & $Pmc2_1$   & 8  & 70.9 & 49.6  & 2.53 & 3.23 & 0.62 & 3.62 & * \\
d-WS & ZZN-63   & 63  & $Cmcm$     & 16 & 56.5 & 56.2  & 1.85 & 2.39 & 2.81 & 4.16 &  \\
WZ   & ZZN-33   & 33  & $Pna2_1$   & 16 & 79.1 & 57.5  & 2.99 & 3.3  & 1.97 & 7.96 &   \\
LC & ZZN-31   & 31  & $Pmn2_1$   & 8  & 60.8 & 60.6  & 1.62 & 1.62 & 0.58 & 0.99 &   \\
d-RS & ZZN-109  & 109 & $I4_1md$   & 16 & 68.1 & 70.8  & 1.29 & 1.96 & 0.85 & 2.13 &   \\
d-RS & ZZN-59   & 59  & $Pmmn$     & 8  & 75.5 & 74.8  & 1.2  & 1.2  & 0.59 & 1.91 &   \\
RS   & ZZN-141  & 141 & $I4_1/amd$ & 16 & 74.9 & 80.7  & 1.15 & 1.87 & 0.83 & 1.96 & * \\
ZB   & ZZN-115  & 115 & $P\bar{4}m2$    & 4  & 102.6  & 85.7  & 2.04 & 3.03 & 0.52 & 1.59 & * \\
d-HX & ZZN-4    & 4   & $P2_1$     & 8  & 77.4   & 89    & 2.38 & 2.73 & 3.75 & 2.28 &   \\
ZB   & ZZN-122  & 122 & $I\bar{4}2d$    & 16 & 111.5  & 92.9  & 2.32 & 2.79 & 0.71 & 7.22 &   \\
RS   & ZZN-12b  & 12 & $C2/m$    & 16 & 100.2  & 99.5  & 1.07 & 1.07 & 0.79 & 2.43 &    \\
d-WS & ZZN-160b & 1   & $P1$       & 24 & 91.9   & 90    & 1.67 & 1.83 & 1.12 & 4.72 &   \\
RS   & ZZN-13   & 166 & $R\bar{3}m$     & 12 & 113.5  & 111.1 & 1.67 & 2.19 & 1.63 & 3.45 &   \\
BN   & ZZN-187   & 187  & $P\bar{6}m2$     & 4  & 113.2   & 113.0  & 0.42 & 0.42 & 1.98 & 2.20 &   \\
RS   & ZZN-123  & 123 & $P4/mmm$   & 4  & 144    & 155.9 & 0    & --   & --   & --   &   \\ \bottomrule
\end{tabular}
}
\end{table*}

\subsection{Grouping structures by prototype}

An important aspect of the analysis in this work involves assigning \ce{ZnZrN2} structures from KNL and the polymorph sampler into a particular prototype category. This is nontrivial, particularly for ternary materials, for the following reasons:

\begin{enumerate}
    \item Simply matching space groups is not sufficient; introducing disorder and distortions often changes the space group.
    \item Some structure classes are commensurate to one another through distortions and lattice transformations (e.g. WZ and BN); how, then, do you categorize intermediate structures?
    \item Some structures have similar short-range motifs but different long-range motifs.
    \item In distorted ternary structures, local symmetry changes and bonds can break such that a structure derived from a simple structure prototype (e.g. RS) no longer appears to look like RS (e.g. ZZN-7, which has 5-fold coordination).
\end{enumerate}

\begin{figure}[!htb]
    \centering
    \includegraphics[width=85mm]{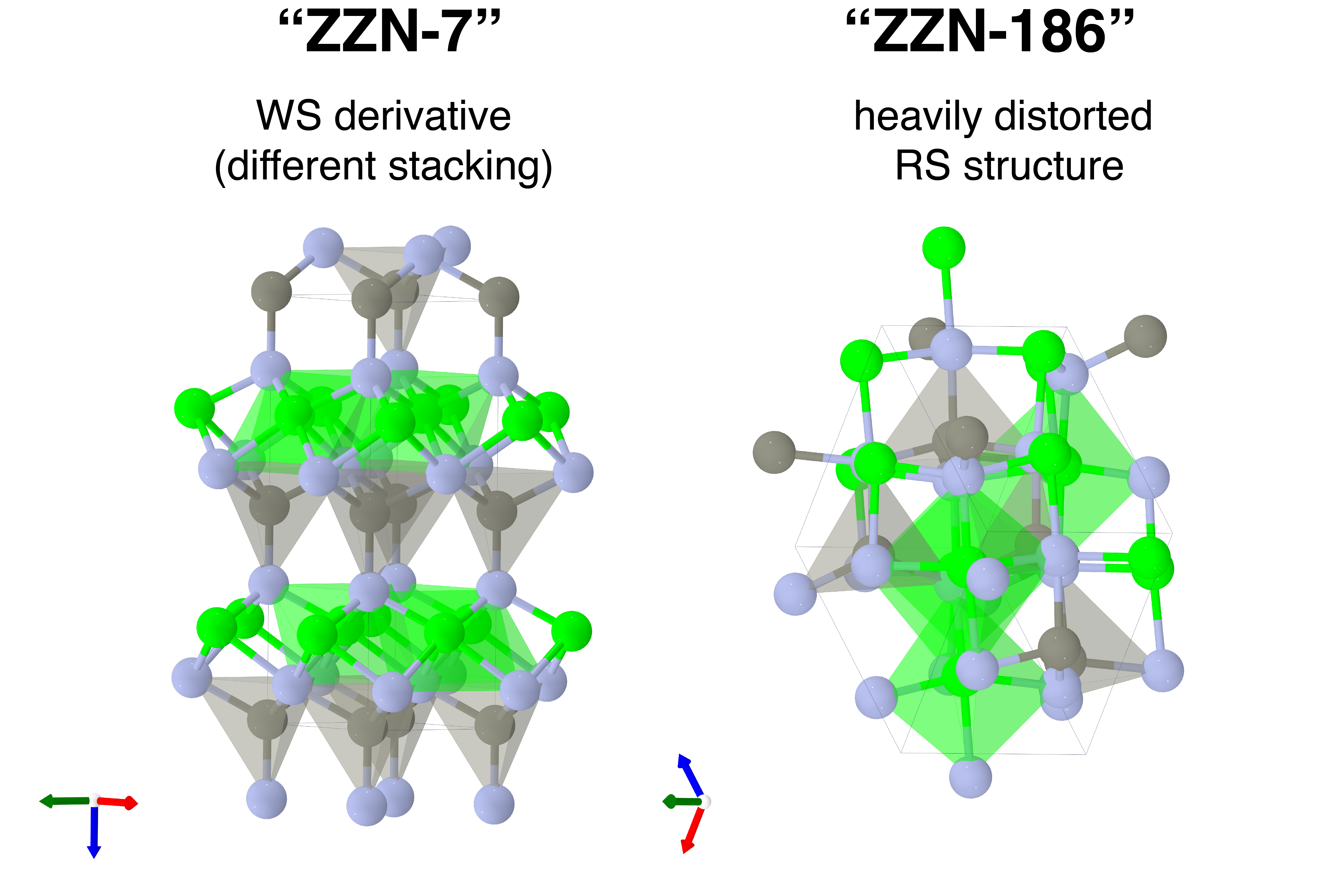}
    \caption{Structures denoted with asterisks in Table \ref{tab-si:prototype_matching}, and their appropriate groupings.}
    \label{fig-si:asterisk-strucs}
\end{figure}

There are multiple methodologies in the literature to classify crystal structures. We initially tried the AFLOW-XtalFinder prototype matcher,\cite{hicks2021aflow} as implemented in \texttt{pymatgen} as \texttt{AflowPrototypeMatcher}, but only half of the structures were assigned matches and over half of those were incorrect (e.g. ZZN-29 was assigned to "SC16 CuCl", and ZZN-8 and ZZN-63 were assigned to "B16" --- upon inspection, these structures do not appear to match). Thus, in this study, to overcome the challenges listed above, we adhere to the following guidelines for structure classification:

\begin{enumerate}
    \item To check structure type, we substitute all cations with Zr to create binary ZrN structures, and relax the structures. Doing so results in the following prototype structures: RS, BN, WZ, ZB, LC, and WS. We select one representative structure from each prototype.
    
    \item For each calculated \ce{ZnZrN2} structure, I structure match to the binary ZrN prototypes
    using the \texttt{pymatgen} \texttt{StructureMatcher} method with the following tolerances and parameters: \texttt{ltol} = 0.5, \texttt{stol} = 0.5, \texttt{angle\_tol} = 10, \texttt{attempt\_supercell} = True, \texttt{allow\_subset} = True.
    
    \item For each structure that matches a given prototype, we calculate the average root mean square (rms) distance between structures, as reported in \textbf{Table \ref{tab-si:prototype_matching}} (see structures in \textbf{Figure \ref{fig-si:asterisk-strucs}}).
    
    \item A structure is then assigned to a given prototype with the smallest rms distance, as reported in "Closest prototype" column.
    
    \item If the closest prototype match has an rms distance greater than 0.08 \AA  (a somewhat arbitrary cutoff motivated by RS, given by \texttt{TRUE} or \texttt{FALSE}), it is classified as distorted with "d-". Note that distorted BN and WZ structures are both classified as "d-HX", since these structures transform into one another through distortions.
    
    \item Structures are inspected individually to check that the assignment is reasonable. Two exceptions, indicated with an * in Table \ref{tab-si:prototype_matching}, are ZZN-186 --- a WS structure with a different stacking rather than d-WS --- and ZZN-7, which matches to both d-WS and RS, but upon inspection appears closer to RS. These structures are depicted in \textbf{Figure \ref{fig-si:asterisk-strucs}} for clarity.
    
    \item The final assigned structure is listed in the column "Prototype assignment," and is used throughout this manuscript.

\end{enumerate}

\begin{table*}[]
\centering
\caption{Prototype matching calculations}
\label{tab-si:prototype_matching}
\resizebox{0.75\textwidth}{!}{
\begin{tabular}{@{}c|ccccccc|ccc|c@{}}
\toprule
  &
  \multicolumn{7}{c|}{\textbf{Structure matcher rms distance}} &
  \multicolumn{3}{c|}{\textbf{Minimum rms distance}} &
    \\ \cmidrule(lr){2-11}
\begin{tabular}[c]{@{}c@{}}\textbf{Structure} \\ \textbf{name}\end{tabular} &
  RS &
  BN &
  WZ &
  ZB &
  WS &
  d-WS &
  LC &
  \begin{tabular}[c]{@{}c@{}}Closest \\ prototype\end{tabular} &
  \begin{tabular}[c]{@{}c@{}}Minimum \\ distance\end{tabular} &
  \begin{tabular}[c]{@{}c@{}}"distorted"? \\ (rms > 0.08)\end{tabular} &
  \begin{tabular}[c]{@{}c@{}}\textbf{Prototype} \\ \textbf{assignment}\end{tabular} \\ \midrule
ZZN-109  & 0.088 & 0.285 & 0.205 & 0.403 & 0.328 & 0.259     & 0.465 & RS    & 0.088 & TRUE  & d-RS  \\
ZZN-115  & 0.261 &   --    &   --    & 0.000 & 0.434 & 0.377     & 0.408 & ZB    & 0.000 & FALSE & ZB    \\
ZZN-12   & 0.193 & 0.076 & 0.136 & 0.324 & 0.424 & 0.390     & 0.354 & BN    & 0.076 & FALSE & BN    \\
ZZN-122  & 0.261 &   --    &   --    & 0.000 & 0.434 & 0.377     & 0.425 & ZB    & 0.000 & FALSE & ZB    \\
ZZN-123  & 0.000 &   --    & 0.390 & 0.256 & 0.398 & 0.235     & 0.445 & RS    & 0.000 & FALSE & RS    \\
ZZN-129  & 0.192 & 0.172 &   --    & 0.442 &   --    & -- & 0.223 & BN    & 0.172 & TRUE  & d-HX  \\


ZZN-13   & 0.000 &   --    & 0.384 & 0.256 & 0.385 & 0.231     & 0.451 & RS    & 0.000 & FALSE & RS    \\
ZZN-14   & 0.359 & 0.275 & 0.156 & 0.324 & 0.448 & 0.209     &   --    & WZ    & 0.156 & TRUE  & d-HX  \\
ZZN-141  & 0.000 & 0.221 & 0.388 & 0.256 & 0.400 & 0.206     & 0.319 & RS    & 0.000 & FALSE & RS    \\
ZZN-156  & 0.422 & 0.339 & 0.377 &   --    & 0.034 & 0.244     & 0.443 & WS    & 0.034 & FALSE & WS    \\
ZZN-160a & 0.250 & 0.437 & 0.392 & 0.446 & 0.240 & 0.115     &   --    & d-WS  & 0.115 & TRUE  & d-WS  \\
ZZN-160b & 0.259 & 0.364 & 0.408 & 0.345 & 0.443 & 0.114     & 0.393 & d-WS  & 0.114 & TRUE  & d-WS  \\
ZZN-164  & 0.401 & 0.387 & 0.435 & 0.442 & 0.071 & 0.248     &   --    & WS    & 0.071 & FALSE & WS    \\
ZZN-186  & 0.382 & 0.306 & 0.376 & 0.461 & 0.287 & 0.348     &   --    & WS    & 0.287 & TRUE  & WS*    \\
ZZN-26   & 0.393 & 0.114 & 0.028 &   --    & 0.362 & 0.406     & 0.377 & WZ    & 0.028 & FALSE & WZ    \\
ZZN-29   & 0.283 & 0.334 & 0.373 &    --   & 0.451 &           & 0.069 & LC    & 0.069 & FALSE & LC    \\
ZZN-31   & 0.412 & 0.209 & 0.225 & 0.435 & 0.405 &     --      & 0.227 & BN    & 0.209 & TRUE  & LC*  \\
ZZN-33   & 0.395 & 0.111 & 0.016 &   --    & 0.375 & 0.411     & 0.382 & WZ    & 0.016 & FALSE & WZ    \\
ZZN-4    & 0.473 & 0.297 &   --    & 0.461 &   --    & 0.369     & 0.419 & BN    & 0.297 & TRUE  & d-HX  \\
ZZN-45   & 0.455 & 0.293 & 0.403 & 0.443 & 0.457 & 0.356     & 0.072 & LC    & 0.072 & FALSE & LC    \\
ZZN-51   &   --    & 0.005 & 0.105 &   --    & 0.345 & 0.333     & 0.325 & BN    & 0.005 & FALSE & BN    \\
ZZN-59   & 0.095 & 0.335 & 0.340 & 0.445 &   --    &      --     & 0.457 & RS    & 0.095 & TRUE  & d-RS  \\
ZZN-63   & 0.319 &   --    &    --   &    --   & 0.383 & 0.026     & 0.362 & d-WS  & 0.026 & FALSE & d-WS  \\
ZZN-7    & 0.187 & 0.360 & 0.287 & 0.357 & 0.436 & 0.184     &    --   & d-WS/RS & 0.184 & TRUE  & d-RS* \\
ZZN-8    & 0.197 & 0.086 & 0.104 & 0.304 & 0.405 & 0.402     & 0.353 & BN    & 0.086 & TRUE  & BN  \\
ZZN-8a   & 0.339 & 0.379 & 0.383 & 0.380 & 0.203 & 0.367     &   --    & WS    & 0.203 & TRUE  & d-WS  \\ \bottomrule
\end{tabular}
}
\end{table*}

\section{Computational methods}



\subsection{SQS calculations}

\subsubsection{Methodology}

We have incorporated some of the \texttt{ATAT} mcsqs tools into \texttt{pymatgen} using the following modules:

\begin{itemize}
    \item \url{pymatgen/command\_line/mcsqs\_caller}
    \item \url{pymatgen/io/atat}
    \item \url{pymatgen/transformations/advanced\_transformations}
\end{itemize}

\noindent We have also added appropriate tests for each of these subpackages. Here, we describe some of the added functionality and methods, which can be applied to other systems beyond \ce{ZnZrN2}.

The class \texttt{Mcsqs} in \url{pymatgen/io/atat} handles inputs and outputs for the crystal definition format used by mcsqs and other \texttt{ATAT} codes. We have updated this method to be compatible with disordered inputs, as well as inputs with spin and oxidation states. \texttt{mcsqs\_caller} defines a \texttt{NamedTuple} class \texttt{Sqs} as the return type for \texttt{run\_mcsqs}, containing the following data:

\begin{itemize}
    \item \texttt{bestsqs}: the SQS structure with the lowest objective function.
    \item \texttt{objective\_function}: the corresponding objective function for the bestsqs structure.
    \item \texttt{allsqs}: a list of all of the sqs structures that are calculated.
    \item \texttt{clusters}: a list of the clusters generated by \texttt{clusters.out}.
    \item \texttt{directory}: the directory where the calculations are stored.
\end{itemize}

\noindent We have added functionalities to run calculations in parallel (\texttt{instances}), and a functionality to detect the number of available cores (this determines the default value of \texttt{instances}). In this study we run on 64 cores on the NERSC supercomputer Cori so that multiple SQS structures can be calculated in parallel; the outputs of each of these 64 calculations are what is fed to the \texttt{allsqs} parameter, and the output with the lowest objective function is fed to the \texttt{bestsqs} parameter. For \texttt{run\_mcsqs}, a \texttt{search\_time} must be specified. We run a series of convergence tests on the RS and WZ structures to ensure that the lowest objective functions are achieved; for small supercells (32 atoms) 4 hours are appropriate, but for large supercells (64 and 128 atoms) we run calculations for 24 hours to ensure that the lowest objective function is reached. Other input parameters we have added are \texttt{temperature} (Monte Carlo temperature, \texttt{T} in ATAT), \texttt{wr} (weight assigned to range of perfect correlation match in objective), \texttt{wn} (multiplicative decrease in weight per additional point in cluster), \texttt{wd} (exponent of decay in weight as function of cluster diameter), and \texttt{tol} (tolerance for matching correlations).

Within \url{pymatgen/transformations/advanced\_transformations}, \texttt{SQSTransformation} is a transformation class that creates a special quasirandom structure (SQS) from a structure with partial occupancies by running mcsqs\_caller.  This class contains static method \texttt{\_sqs\_cluster\_estimate}, which set up an \texttt{ATAT} cluster.out file for a given structure and set of constraints for a given variable \\ \texttt{cluster\_size\_and\_shell}, which assigns the maximum number of shells to include for a given cluster size (e.g. the default value \texttt{\{2:3, 3:2, 4:1\}} would include 3 shells for doublets, 2 shells for triplets, and 1 shell for quadruplets). We have adopted this from the literature, and use this default value for our calculations, but acknowledge the limitations below. This class also contains \texttt{\_get\_unique\_bestsqs\_strucs}, which structure matches the output SQS structures and returns only the unique structures.

Finally, to run the transformation we use the \texttt{apply\_transformation} function, which is compatible with other \texttt{pymatgen} transformations and returns either the \texttt{bestsqs} structure or, if \texttt{return\_ranked\_list = TRUE}, returns all of the sqs structures ("ranked" by objective function). We point the reader to \texttt{pymatgen} docs for more information: \url{https://pymatgen.org/pymatgen.command\_line.mcsqs\_caller.html#pymatgen.command\_line.mcsqs\_caller.Sqs}.

\subsubsection{Calculated \ce{ZnZrN2} SQS structures}

For each ZrN structure prototype listed above and in Tables \ref{tab-si:SI_calcs} and \ref{tab-si:prototype_matching} --- RS, BN, WZ, ZB, WS, LC, d-RS, d-HX, d-WS --- we select a representative structure to use as input structures for SQS calculations. For each input, we create a "disordered structure," i.e. a structure with partial occupancies on the cation side of 50\% Zn and 50\% Zr such that the stoichiometry adds up to \ce{ZnZrN2}. An example of what these disordered structures looks like for WS and RS is shown in Figure 5 in the main text.

Then, for each of these nine disordered structures, we apply \texttt{SQSTransformation} with the following inputs:

\begin{itemize}
    \item Super cell sizes of 16, 32, 64, 128
    \item Run times of 4 hours and 24 hours (see above; 16 atom super cells typically require only 30 minutes to converge)
    \item \texttt{cluster\_size\_and\_shell} = \{2:3, 3:2, 4:1\}
    \item Default parameters: \texttt{temperature}=1, \texttt{wr}=1, \texttt{wn}=1, \texttt{wd}=0.5, \texttt{tol}=\num{1e-3}
\end{itemize}

For each supercell size, we limit our calculations by selecting 4 output structures with the lowest objective functions (note that in some cases there are only 2 or 3 unique structures, so in those cases we use only 2 or 3 output structures). This results in 16 SQS structures for each prototype, and thus 144 SQS structures total. We then relax each of these 144 structures using PBE to calculate formation enthalpy, with the methods described above. Many of the 128 atom supercell structures do not converge properly, so the 128 atom supercells are excluded from our analysis.

We also check whether each SQS structure has transformed into another structure type after relaxation, using the same approach outlined in the "Grouping structures by prototype" subsection. We restrict our set of SQS structures to the 64 atom structures for the manuscript, since 16 and 32 atoms is likely too small to approximate "random" disorder (but we calculate the structures anyways to ensure that their orderings are not lower energy than our ordered structures) and some of the 128 atom cells were too large to converge within a reasonable time frame. We rerun the 64 atom cell SQS structures using the SCAN functional to more accurately estimate the formation energies. And lastly, for a given prototype class, we average the resulting SQS SCAN formation energies to calculate the "random disorder polymorph energy" ($E_\mathrm{rdp}$) reported in Figure 5 in the manuscript.

\subsubsection{Limitations of SQS}

SQS is used here as an approximate for a fully disordered structure. For each prototype structure, the SQS with the lowest formation enthalpy is used to determine $E_\mathrm{rdp}$. It is important to acknowledge the limitations of the SQS approach in modeling disordered structures and calculating $E_\mathrm{rdp}$:

\begin{enumerate}
    \item \textbf{Neglecting short-range order effects:} In this case, and almost always, SQS is used for fully random structures without including short range order (SRO) effects. We have assumed an arbitrary set of clusters and weights, as described above. The SQS could in principle be fitted and clusters could be defined to reproduce SRO structures, but that is very rarely done, since one must know the correlation functions for the SRO in the first place; if this is done (e.g. via cluster expansion Monte Carlo, CE-MC), the SQS is usually not needed any more (although it can be useful for post-processing, e.g. very expensive electronic structure calculations like GW). We cannot \textit{a priori} know what the SRO effects are for each \ce{ZnZrN2} prototype, so this is a limitation here.
    \item \textbf{Selection bias:} We have assumed that for each structure type, the "bestsqs" structures will have the lowest formation enthalpies. However, since the objective function is determined by the clusters, which are somewhat arbitrarily selected and neglect SRO, we could be optimizing around the wrong local minima and missing the actual lowest energy disordered structures.
    \item \textbf{Incomplete data set:} Similarly, since for each prototype class we only select the four unique structures with the lowest objective functions, we could be neglecting structures that actually have lower formation energy. As in most computational approaches, there is a trade-off between computational cost (more structures) and accuracy.
\end{enumerate}

\noindent For these reasons, our $E_\mathrm{rdp}$ can be considered an upper limit of the gain in configurational enthalpy upon disorder.

\subsubsection{SQS calculations of all prototype structure classes}

\begin{figure}[!htb]
    \centering
    \includegraphics[width=85mm]{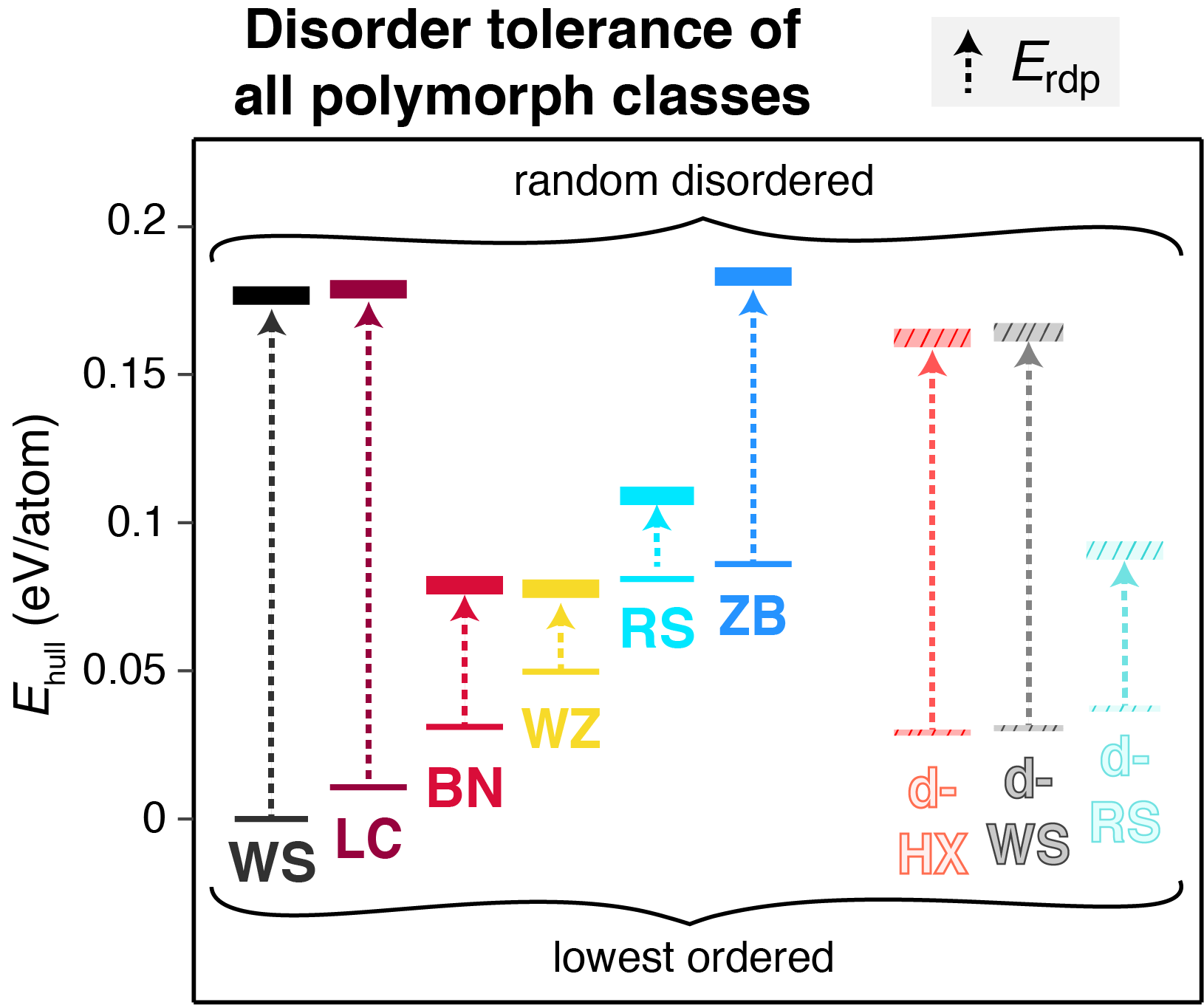}
    \caption{An extension of Figure 5(b) in the manuscript to include all of the prototype structure classes}.
    \label{fig-si:si-sqs}
\end{figure}

\textbf{Figure \ref{fig-si:si-sqs}} plots the SQS formation energies for all of the structure prototype classes. We observe that WZ, similar to BN and RS, has a low random disordered formation energy ($\Delta H_{k}^\mathrm{dis}$) and thus a small $E_{\mathrm{rdp}}$. In contrast, the $\Delta H_{k}^\mathrm{dis}$ for LC, ZB, and d-HX jumps to over 0.150 eV/atom, on par with that of WS and d-WS. This is reasonable according to our disorder tolerance hypothesis, since the WZ prototype structure has equivalent cation sites and the LC prototype structure has in-equivalent cation sites. The explanation for ZB and d-HX is somewhat more nuanced, since the prototypes for both of these classes have equivalent cation sites. Upon inspection, the SQS structures for d-HX are so distorted that the cation sites are only equivalent for a very loose tolerance, so this increased distortion could be contributing to the increase in configurational enthalpy upon disorder. However, this does not appear to be the case for the ZB SQS structures, and so the explanation for such a high $\Delta H_{k}^\mathrm{dis}$ in unknown.

\subsection{Polymorph sampler}

\subsubsection{Additional polymorph sampler details}

Previous research using the polymorph sampler statistical method showed that crystalline structures with high ensemble probabilities --- i.e. high occurrence in random sampling and low formation energies --- correlated with synthetically realized polymorphs in the chemical systems of MgO, ZnO, \ce{SnO2}, and Si, among others.\cite{stevanovic2016sampling, jones2017polymorphism, jones2020glassy} Experimental enthalpies, powder XRD, and radial distribution functions of amorphous Si were also all reproduced with high accuracy when compared to the corresponding computed ensemble-averaged quantities.

In our assessment of the \ce{ZnZrN2} system using the polymorph sampler, due the presence of 5,000 structures the above structure matching procedure is not tractable and an alternate classification scheme is adopted. Originally, the classification of structures obtained by the \textit{ab-initio} random structure sampling into classes of equivalent structures was based on the equality of the total energy, volume, space group assignment, and coordination up to the third shell. Because of the cation disorder in \ce{ZnZrN2}, the structures that differ only in the cation distribution, but not in the underlying lattice, may differ in all of the above criteria. To account for these effects, the classification is based on the underlying lattice rather than the energy, volume, space group assignment, and coordination. The underlying lattice assignment is carried out by symmetrizing the structure with all cations labeled the same and evaluating the resulting space group. The structures are then grouped into the classes with the same underlying space group and coordination of atoms. Because of the spread in energy present in these classes that is due to differences in the cation distribution, the ensemble probabilities for these structures are evaluated from the partial partition functions as explained in the manuscript.

\subsubsection{XRD models}

The XRD models for RS and BN, reported in Figure 2(c) in the manuscript, are calculated from the polymorph sampler structures. Essentially, for each prototype space group class, all of the computed powder XRD patterns are averaged across all structures within a class. Classes that have resulted XRD patters with one very diffuse peak centered around 30--35\degree in 2$\theta$ represent amorphous structure classes, which have been excluded from the analysis due to their high formation energies. Classes that have ``sharp'' average XRD patterns clearly represent a given disordered structure. In this model, some specific peaks of ordered structures are essentially smoothed out, and the remaining peaks correspond to the shared peaks that would be observed in a disordered structure. In our case, the disordered RS and disordered BN models from this method correspond very well to what is observed experimentally. Additionally, this classification based on XRD gives another "check" to the polymorph sampler structure classification scheme addressed above.

We note that since the RS and BN standards are ensemble averages of multiple XRD patterns, representing disordered structures, it is not trivial to link a given observed peak to a reflection index. The reflection index varies by ordered prototype, and no single prototype structure gives the pattern that the disordered aggregate model reveals. For example, in BN the highest experimental BN reflection at $Q = $ 2.45 \AA \space appears to be a convolution of multiple reflections. For a single ordered structure, this reflection corresponds to (121), (221), (023), (201) and/or (003). Considering other ordered structures and SQS structures, there are even more corresponding indices. Thus, for an approximation to label peaks in Figure 2 in the manuscript and for simplicity, we refer to the reflections of a single simple binary BN structure: (002), (100), (101), (102), etc. Other index notations are likely for larger supercell structures of simple hBN structure, because of the presence of Zn and Zr.


\subsection{Vibrational contributions to free energy}


To estimate energetic contributions from vibrational degrees of freedom for structures of interest, density functional perturbation theory (DFPT) calculations for gamma ($\Gamma$) point phonons ($q$ = 0) are run on representative polymorphs. These calculations are performed on ordered \ce{ZnZrN2} polymorphs of RS, BN, and WS structures with spacegroup numbers 141 ($I4_1/amd$), 8 ($Cm$), and 156 ($P3m1$), respectively, since these polymorphs have the lowest $E_\mathrm{hull}$ for each of the structure types (see Table 1). First a highly optimized GGA+U structure relaxation is performed for each of these polymorphs, with settings consistent with GGA+U calculations in the rest of the paper (Zr U = 3 eV/atom). Second, a DFPT calculation (\texttt{IBRION}=8 in VASP) is performed on the relaxed unit cells of each polymorph. Third, the \texttt{phonopy} package is used for post-processing to compute thermal properties, including the vibrational contributions to the free energy.\cite{togo2015phonopy} A series of convergence tests are performed to ensure results converge with respect to structure optimization. We refer the reader elsewhere for details about this method.\cite{gonze1997dynamical}

The Gibbs free energy $G$ of a system is defined by:

\begin{equation}
    G(T) = U - TS + PV
    \label{eq:gibbs-free-energy}
\end{equation}

\noindent where $U$ is the internal energy, $T$ the temperature, $S$ the entropy, $P$ the pressure, and $V$ the volume. Using DFPT to include vibrational degrees of freedom, we approximate the free energy $G_k^\mathrm{vib}$ of a given polymorph $k$ as the following: 

\begin{equation}
    G_{k}^\mathrm{vib}(T) = \Delta H_{k} + U_{k}^\mathrm{vib} - TS_{k}^\mathrm{vib}
    \label{eq:vib-energy}
\end{equation}

\noindent where $\Delta H$ is the enthalpy of formation, $U^\mathrm{vib}$ is the vibrational contribution to the internal energy, and $S^\mathrm{vib}$ is the vibrational entropy (note that the $PV$ term is excluded here). Both $U^\mathrm{vib}$ and $S^\mathrm{vib}$ are computed using \texttt{phonopy}.\cite{togo2015phonopy} For ease of comparison with the results in the manuscript, we then reference the free energy to the ground state $E_\mathrm{hull}$ (see Figure 1) for each polymorph $k$ such that:

\begin{equation}
    G_{k}^\mathrm{vib}(T) = E_{\mathrm{hull},k} + U_{k}^\mathrm{vib} - TS_{k}^\mathrm{vib}
    \label{eq:vib-norm}
\end{equation}

\noindent We note the distinction between this free energy $G_{k}^\mathrm{vib}$ which is a function of actual temperature $T$ and the configurational free energy $G_{k}^\mathrm{conf}$ from Figure 3(b) (Equation 2) which is a function of the effective temperature $T_\mathrm{eff}$ (a proxy for disorder), and therefore these two equations cannot be simply combined.

\begin{figure}
    \centering
    \includegraphics[width=100mm]{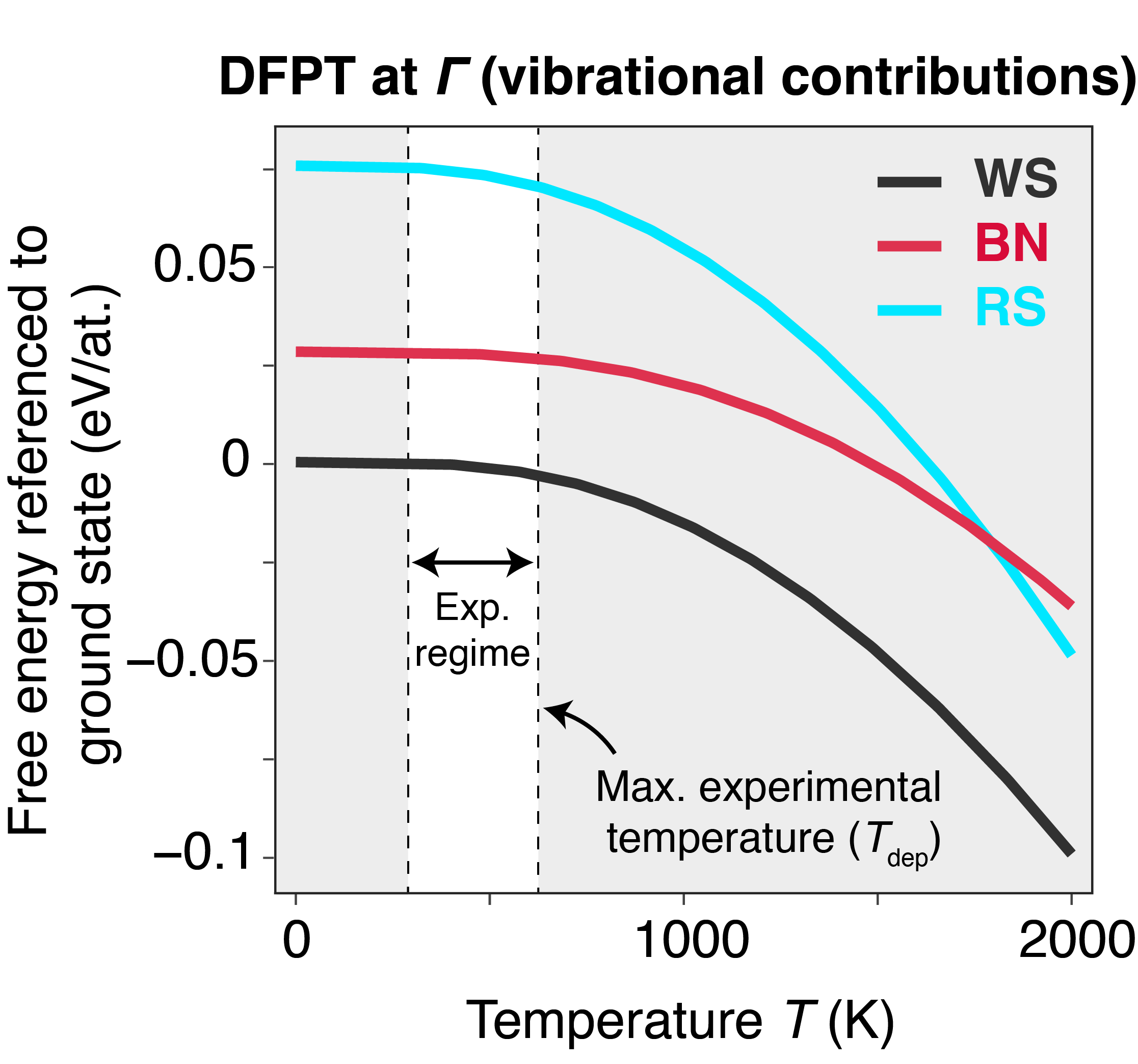}
    \caption{Free energies including vibrational contributions for three representative phases of WS, BN, and RS, computed from DFPT for phonons at the gamma point ($G_{k}^\mathrm{vib}$ from \autoref{eq:vib-norm}). Free energies are referenced to the ground state at 0 K and are plotted as a function of temperature. The vertical dotted lines correspond to the minimum and maximum temperatures achieved experimentally ($\sim$65 \degree C or $\sim$338 K and $\sim$350 \degree C or $\sim$623 K, respectively; see Figure 2b), with the regions below and above these temperatures in grey to indicate they were likely not probed in this experiment.}
    \label{fig-si:si-vib}
\end{figure}

\textbf{\autoref{fig-si:si-vib}} plots the resulting free energies $G_{k}^\mathrm{vib}(T)$ from DFPT, referenced to the ground state at 0 K and as a function of temperature from 0--2000 K. These results show that RS is somewhat destabilized with respect to BN at very high temperatures ($\sim$1800 K), but across all temperatures WS is still the lowest energy structure compared to BN or RS. Furthermore, energy ordering shifts only negligibly within the range of temperatures probed experimentally (65\degree C $\leq T_\mathrm{dep} \leq$ 350\degree C). Therefore, vibrational effects to not explain the stabilization of BN and RS over the WS phase observed in our experiments. Rather, by comparing $G_{k}^\mathrm{vib}$ to $G_{k}^\mathrm{conf}$ as done in Figure 4 in the main text, we show RS and BN are stabilized and WS is destabilized at high temperature by \textit{configurational} degrees of freedom rather than vibrational degrees of freedom. We comment that the DFPT calculations here serve to \textit{estimate} the vibrational contributions to free energy --- a more rigorous analysis would consider all polymorphs, probe other values of $q$, and compute anharmonic contributions --- however such an analysis is computationally expensive and is beyond our scope. Here, DFPT is sufficient to rule out any major energetic reordering due to vibrational entropy.

\subsection{Alloy energy estimates}

\subsubsection{Alloy structure generation}

To estimate formation energies for alloy structures, we calculate structures along the tie-line \ce{Zn_{x}Zr_{1-x}N} as reported in Figure 4. As noted in the manuscript, the experimental composition likely deviates from this tie-line, but it is the simplest theoretical model to study. We limit calculations to $x$ = 0, 0.25, 0.5, and 0.75; $x$ = 1 is unphysical due to nominal valence, and including these results could be misleading (we note that $x$ = 0.75 is likely on the limit of the unphysical regime).

For $x$ = 0, i.e. \ce{ZrN}, we start with each of the \ce{ZnZrN2} polymorphs and substitute all Zn atoms for Zr. Then the structures are relaxed using PBE. We note that several of the polymorph structures transform into another structure type upon relaxation. We assess this by reporting the space group, and then matching each structure type according the structure matching procedure outlined above. For $x$ = 0, are the structures that are used as prototype inputs for structure matching in the "Grouping structures by prototype" section above. Note that WZ is not present at $x$ = 0, as it has relaxed into BN, d-RS has relaxed into RS, and d-HX has relaxed in RS or BN. $x$ = 0 calculations are shown here, but are excluded from the manuscript.

For alloy structures $x$ = 0.25 and $x$ = 0.75, we start with the $x$ = 0.5 polymorphs as initial structures. Supercells must contain at least 8 atoms (and contain a total number of atoms divisible by 8) in order to satisfy a 1:3 ratio of Zr:Zn or Zn:Zr. To satisfy this constraint, for each polymorph, we check whether the number of atoms is divisible by 8 and if not, create the smallest possible supercell that is. The we create a disordered structure, and apply an \texttt{OrderDisorderedStructureTransformation} to select four possible orderings (ranked by Ewald sum). For each ordering, we perform a PBE structure relaxation, followed by SCAN calculations which have been reported in the manuscript. We then check whether structures have transformed into a new structure class using the procedure outlined previously.

We note that with all of the $x$ = 0, 0.25, 0.5, and 0.75 calculations, we apply the Materials Project (MP) correction scheme to compare energies and create a binary hull phase diagram.\cite{mp2021documentation} This is done by creating \texttt{ComputedEntry} classes for each compound, with their computed energies, associated corrections from the \texttt{MPCompatibility} class, and other structural and identifying information. We also note that this is an approximation based on \textit{ordered} structures, and that energetic orderings may be different when considering disordered structures.

\begin{figure}
    \centering
    \includegraphics[width=70 mm]{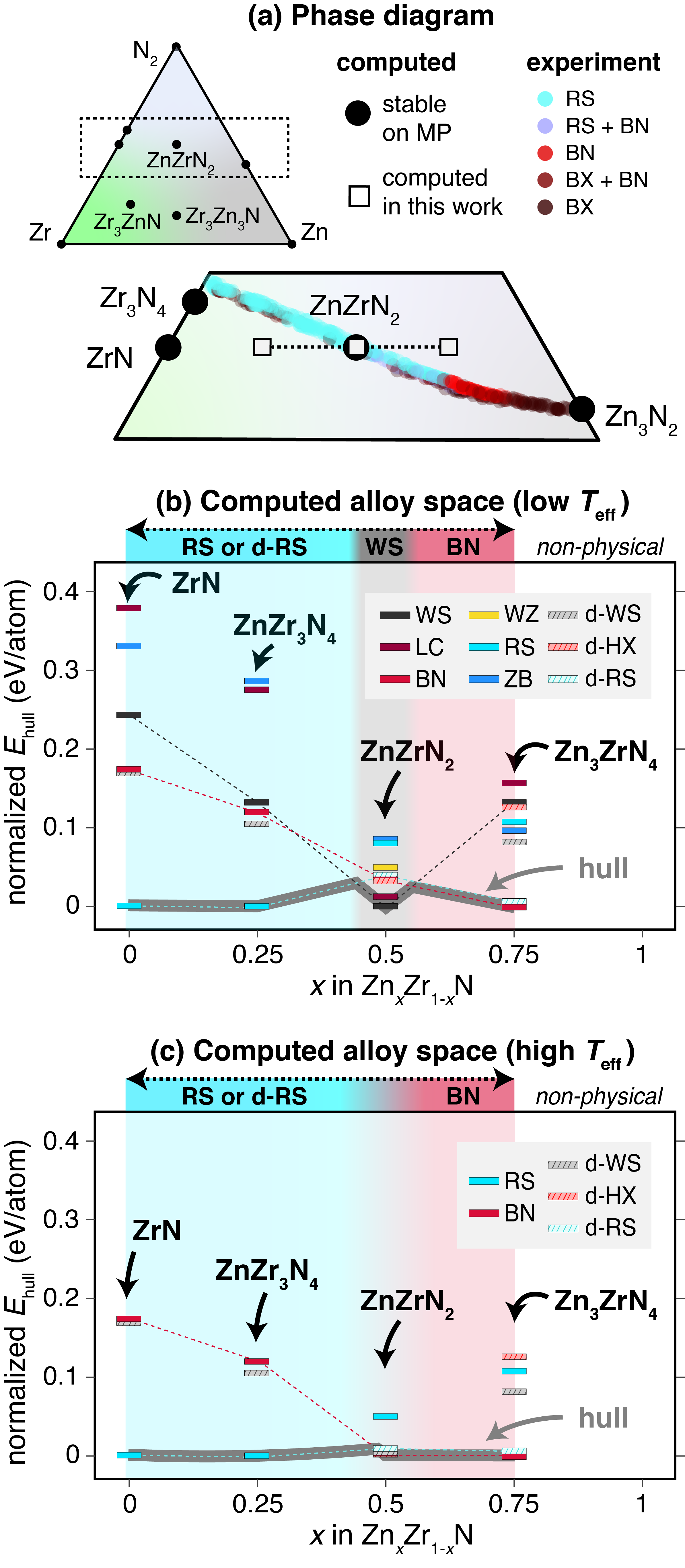}
    \caption{(a) Ternary phase diagram of Zn-Zr-N as reported in the manuscript. (b) Computed phases across tieline ZrN--\ce{Zn3ZrN4}. For each structure class, the lowest formation energy structure is plotted, and dotted lines connect such values (note absent classes: d-RS and d-HX relax into RS at $x$ = 0, 0.25; WZ relaxes into BN at $x \neq$ 0.5). Formation energies are referenced to the lowest energy structure at each composition "referenced $E_\mathrm{hull}$"), and the approximate hull is labeled. Background shading and the colored bar represent the lowest energy phase on the hull (gradient accounts for uncertainties in this model). (c) The same space as (b), but including only phases that emerged from the polymorph sampler with low energies (high probabilities) at high effective temperatures ($T_\mathrm{eff}$).}
    \label{fig-si:si-alloy}
\end{figure}

\subsubsection{Alloy hull plots}

\textbf{\autoref{fig-si:si-alloy}}(a) plots ternary Zn-Zr-N phase space, as in the manuscript. \autoref{fig-si:si-alloy}(b) approximates how relative polymorph stability changes with $x$, and indeed shows a narrow window of $x$ in which WS has the lowest $E_\mathrm{hull}$ (grey shading), suggestive of a line compound. RS is the lowest energy polymorph for a wide window of approximately 0 < $x \lesssim$ 0.45 (blue shading) and BN is the lowest energy polymorph at approximately $x \gtrsim$ 0.55 (red shading). Since the LC, ZB, and WZ phases do not emerge in the polymorph sampler and WS is highly metastable at high $T_\mathrm{eff}$ (see Figure 3 in the manuscript), \autoref{fig-si:si-alloy}(c) plots (b) except with these phases removed in order to approximate alloy space at high $T_\mathrm{eff}$. This suggests a phase change from RS to BN somewhere within approximately 0.45 < $x$ < 0.55, which corroborates experimental findings. In (b) and (c) the compositions $x$ > 0.75 (labeled “non-physical”) are not plotted; ZnN is a non-physical unstable compound since Zn is not stable in a +3 oxidation state, yet another phase change to BX at high Zn content is plausible since we observe BX \ce{Zn3N2} experimentally.

\subsection{"\textit{x}log\textit{x}" configurational entropy estimates}

We attempted to estimate configurational entropy $S_\mathrm{conf}$ for the 64 atom SQS random disordered structures. For a structure with just one type of site (i.e. symmetrically equivalent sites), where $x$ is the occupancy of Zn (in this case, $x$ = 0.5), configurational entropy per atom $s_\mathrm{conf}$ is conventionally estimated using the solid solution model of "$x$log$x$" as:

\begin{equation}
    s_\mathrm{conf} = -k_\mathrm{B} [x \mathrm{ln}x + (1-x)\mathrm{ln}(1-x)]
\end{equation}

\noindent and total configurational entropy $S_\mathrm{conf}$ as

\begin{equation}
    S_\mathrm{conf} = N s_\mathrm{conf}
\end{equation}

\noindent where $N$ is the number of cation sites (i.e. 32 for 64 atom SQS cells).

Since $x$ = 0.5, the solution for \ce{ZnZrN2} polymorphs with equivalent sites, namely RS, ZB, WZ, BN, d-RS, and d-HX, are equivalent:

\begin{equation}
    S_\mathrm{conf} = -k_\mathrm{B} (32) [0.5 \mathrm{ln}0.5 + (1-0.5)\mathrm{ln}(1-0.5)] = 1.911 \mathrm{meV/K}
\end{equation}

\begin{figure}[!htb]
    \centering
    \includegraphics[width=\textwidth]{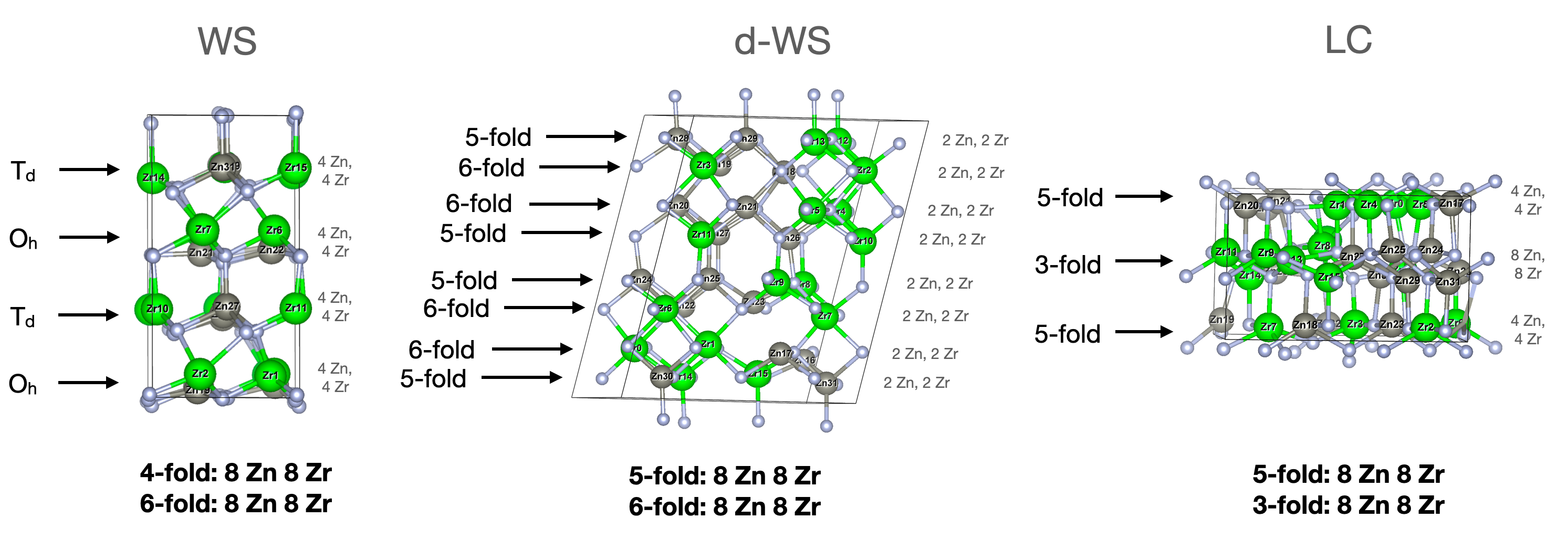}
    \caption{Representative SQS structures with inequivalent cation sites (WS, d-WS, and LC), and counting of cation site occupation for $x$log$x$ configurational entropy estimate.}
    \label{fig-si:xlogx}
\end{figure}

For our structures with symmetrically \textit{inequivalent} sites --- WS, LC, d-WS --- this formula can be generalized such that $\lambda_i$ is the occupancy of Zn on the $i$th site and:

\begin{equation}
    s_{i} = -k_\mathrm{B} [\lambda_i \mathrm{ln}\lambda_i + (1-\lambda_i)\mathrm{ln}(1-\lambda_i)]
\end{equation}

\begin{equation}
    s_\mathrm{conf} = \sum_i s_{i}
\end{equation}

\noindent Then, for each SQS structure, we count the occupancy of Zn and Zr in each site, as depicted for WS, LC, and d-WS in \textbf{Figure \ref{fig-si:xlogx}}. We find that for each structure, $\lambda_i$ = 0.5; using this simple estimate $S_\mathrm{conf}$ is the same as that of symmetrically equivalent sites. Thus, including $S_\mathrm{conf}$ in the free energy of the random disordered structures would just shift the energy of each \ce{ZnZrN2} SQS structure by the same quantity, leaving relative energy orderings unchanged. For this reason we exclude this contribution from the energy estimates in the manuscript.

\pagebreak
\bibliographystyle{ieeetr}
\bibliography{refs.bib}
